\documentclass[conference]{IEEEtran}
\IEEEoverridecommandlockouts

\usepackage{anyfontsize}
\usepackage{amsmath,amsfonts,amssymb,amsthm}
\usepackage{graphicx}
\usepackage{tikz}
\usepackage{hyperref}
\usepackage[backgroundcolor=magenta!10!white]{todonotes}
\usepackage{setspace}
\usepackage[inline]{enumitem}
\usepackage{ifthen}
\usepackage{caption}
\usepackage{microtype}
\usepackage{stackrel}
\usepackage{proof}
\usepackage[space]{cite}
\usepackage{numprint}
\usepackage{cleveref}
\usepackage{multirow}
\usepackage{subcaption}
\usepackage{placeins}
\usepackage{mathpazo}

\usepackage{tikz}
\usetikzlibrary{positioning,arrows,automata,calc,shapes}

\definecolor{darkgreen}{rgb}{0.0, 0.5, 0.0}
\definecolor{darkred}{rgb}{0.9, 0.0, 0.0}
\definecolor{darkblue}{rgb}{0.0, 0.0, 0.9}

\tikzset{
	state/.style={rounded rectangle,draw=black,inner sep=2mm,minimum width=9mm,minimum height=5mm},
	rstate/.style={rectangle,draw=black,inner sep=1ex,minimum width=9mm,minimum height=5mm},
	istate/.style={minimum width=5mm, minimum height=6mm},
	bullet/.style={circle,draw=black,fill=black,inner sep=0cm, minimum size=0.7mm},
	initial text={},
	every initial by arrow/.style={->,>=stealth'},
	ptran/.style={rounded corners, ->,>=stealth',auto},
	ntran/.style={rounded corners, -,auto}
}

    \usepackage{graphicx}
    \makeatletter
    \def\maxwidth{\ifdim\Gin@nat@width>\linewidth\linewidth
    \else\Gin@nat@width\fi}
    \makeatother
    \let\Oldincludegraphics\includegraphics
    \renewcommand{\includegraphics}[1]{\Oldincludegraphics[width=.8\maxwidth]{#1}}

    \usepackage{adjustbox} 
    \usepackage{xcolor} 
    \usepackage{enumerate} 
    \usepackage{amsmath} 
    \usepackage{amssymb} 
    \usepackage{textcomp} 
    \AtBeginDocument{%
    }
    \usepackage{upquote} 
    \usepackage{eurosym} 
    \usepackage[mathletters]{ucs} 
    \usepackage[utf8x]{inputenc} 
    \usepackage{fancyvrb} 
    \fvset{listparameters=\setlength{\topsep}{3pt}\setlength{\partopsep}{0pt}}

    \usepackage{grffile} 
    \usepackage{hyperref}
    \usepackage{longtable} 
    \usepackage{booktabs}  
    \usepackage[inline]{enumitem} 
    \usepackage[normalem]{ulem} 

    \definecolor{urlcolor}{rgb}{0,.145,.698}
    \definecolor{linkcolor}{rgb}{.71,0.21,0.01}
    \definecolor{citecolor}{rgb}{.12,.54,.11}

    \definecolor{ansi-black}{HTML}{3E424D}
    \definecolor{ansi-black-intense}{HTML}{282C36}
    \definecolor{ansi-red}{HTML}{E75C58}
    \definecolor{ansi-red-intense}{HTML}{B22B31}
    \definecolor{ansi-green}{HTML}{00A250}
    \definecolor{ansi-green-intense}{HTML}{007427}
    \definecolor{ansi-yellow}{HTML}{DDB62B}
    \definecolor{ansi-yellow-intense}{HTML}{B27D12}
    \definecolor{ansi-blue}{HTML}{208FFB}
    \definecolor{ansi-blue-intense}{HTML}{0065CA}
    \definecolor{ansi-magenta}{HTML}{D160C4}
    \definecolor{ansi-magenta-intense}{HTML}{A03196}
    \definecolor{ansi-cyan}{HTML}{60C6C8}
    \definecolor{ansi-cyan-intense}{HTML}{258F8F}
    \definecolor{ansi-white}{HTML}{C5C1B4}
    \definecolor{ansi-white-intense}{HTML}{A1A6B2}

    \DefineVerbatimEnvironment{Highlighting}{Verbatim}{%
    commandchars=\\\{\},\topsep=20pt}

\makeatletter
\def\PY@reset{\let\PY@it=\relax \let\PY@bf=\relax%
    \let\PY@ul=\relax \let\PY@tc=\relax%
    \let\PY@bc=\relax \let\PY@ff=\relax}
\def\PY@tok#1{\csname PY@tok@#1\endcsname}
\def\PY@toks#1+{\ifx\relax#1\empty\else%
    \PY@tok{#1}\expandafter\PY@toks\fi}
\def\PY@do#1{\PY@bc{\PY@tc{\PY@ul{%
    \PY@it{\PY@bf{\PY@ff{#1}}}}}}}
\def\PY#1#2{\PY@reset\PY@toks#1+\relax+\PY@do{#2}}

\expandafter\def\csname PY@tok@w\endcsname{\def\PY@tc##1{\textcolor[rgb]{0.73,0.73,0.73}{##1}}}
\expandafter\def\csname PY@tok@c\endcsname{\let\PY@it=\textit\def\PY@tc##1{\textcolor[rgb]{0.25,0.50,0.50}{##1}}}
\expandafter\def\csname PY@tok@cp\endcsname{\def\PY@tc##1{\textcolor[rgb]{0.74,0.48,0.00}{##1}}}
\expandafter\def\csname PY@tok@k\endcsname{\let\PY@bf=\textbf\def\PY@tc##1{\textcolor[rgb]{0.00,0.50,0.00}{##1}}}
\expandafter\def\csname PY@tok@kp\endcsname{\def\PY@tc##1{\textcolor[rgb]{0.00,0.50,0.00}{##1}}}
\expandafter\def\csname PY@tok@kt\endcsname{\def\PY@tc##1{\textcolor[rgb]{0.69,0.00,0.25}{##1}}}
\expandafter\def\csname PY@tok@o\endcsname{\def\PY@tc##1{\textcolor[rgb]{0.40,0.40,0.40}{##1}}}
\expandafter\def\csname PY@tok@ow\endcsname{\let\PY@bf=\textbf\def\PY@tc##1{\textcolor[rgb]{0.67,0.13,1.00}{##1}}}
\expandafter\def\csname PY@tok@nb\endcsname{\def\PY@tc##1{\textcolor[rgb]{0.00,0.50,0.00}{##1}}}
\expandafter\def\csname PY@tok@nf\endcsname{\def\PY@tc##1{\textcolor[rgb]{0.00,0.00,1.00}{##1}}}
\expandafter\def\csname PY@tok@nc\endcsname{\let\PY@bf=\textbf\def\PY@tc##1{\textcolor[rgb]{0.00,0.00,1.00}{##1}}}
\expandafter\def\csname PY@tok@nn\endcsname{\let\PY@bf=\textbf\def\PY@tc##1{\textcolor[rgb]{0.00,0.00,1.00}{##1}}}
\expandafter\def\csname PY@tok@ne\endcsname{\let\PY@bf=\textbf\def\PY@tc##1{\textcolor[rgb]{0.82,0.25,0.23}{##1}}}
\expandafter\def\csname PY@tok@nv\endcsname{\def\PY@tc##1{\textcolor[rgb]{0.10,0.09,0.49}{##1}}}
\expandafter\def\csname PY@tok@no\endcsname{\def\PY@tc##1{\textcolor[rgb]{0.53,0.00,0.00}{##1}}}
\expandafter\def\csname PY@tok@nl\endcsname{\def\PY@tc##1{\textcolor[rgb]{0.63,0.63,0.00}{##1}}}
\expandafter\def\csname PY@tok@ni\endcsname{\let\PY@bf=\textbf\def\PY@tc##1{\textcolor[rgb]{0.60,0.60,0.60}{##1}}}
\expandafter\def\csname PY@tok@na\endcsname{\def\PY@tc##1{\textcolor[rgb]{0.49,0.56,0.16}{##1}}}
\expandafter\def\csname PY@tok@nt\endcsname{\let\PY@bf=\textbf\def\PY@tc##1{\textcolor[rgb]{0.00,0.50,0.00}{##1}}}
\expandafter\def\csname PY@tok@nd\endcsname{\def\PY@tc##1{\textcolor[rgb]{0.67,0.13,1.00}{##1}}}
\expandafter\def\csname PY@tok@s\endcsname{\def\PY@tc##1{\textcolor[rgb]{0.73,0.13,0.13}{##1}}}
\expandafter\def\csname PY@tok@sd\endcsname{\let\PY@it=\textit\def\PY@tc##1{\textcolor[rgb]{0.73,0.13,0.13}{##1}}}
\expandafter\def\csname PY@tok@si\endcsname{\let\PY@bf=\textbf\def\PY@tc##1{\textcolor[rgb]{0.73,0.40,0.53}{##1}}}
\expandafter\def\csname PY@tok@se\endcsname{\let\PY@bf=\textbf\def\PY@tc##1{\textcolor[rgb]{0.73,0.40,0.13}{##1}}}
\expandafter\def\csname PY@tok@sr\endcsname{\def\PY@tc##1{\textcolor[rgb]{0.73,0.40,0.53}{##1}}}
\expandafter\def\csname PY@tok@ss\endcsname{\def\PY@tc##1{\textcolor[rgb]{0.10,0.09,0.49}{##1}}}
\expandafter\def\csname PY@tok@sx\endcsname{\def\PY@tc##1{\textcolor[rgb]{0.00,0.50,0.00}{##1}}}
\expandafter\def\csname PY@tok@m\endcsname{\def\PY@tc##1{\textcolor[rgb]{0.40,0.40,0.40}{##1}}}
\expandafter\def\csname PY@tok@gh\endcsname{\let\PY@bf=\textbf\def\PY@tc##1{\textcolor[rgb]{0.00,0.00,0.50}{##1}}}
\expandafter\def\csname PY@tok@gu\endcsname{\let\PY@bf=\textbf\def\PY@tc##1{\textcolor[rgb]{0.50,0.00,0.50}{##1}}}
\expandafter\def\csname PY@tok@gd\endcsname{\def\PY@tc##1{\textcolor[rgb]{0.63,0.00,0.00}{##1}}}
\expandafter\def\csname PY@tok@gi\endcsname{\def\PY@tc##1{\textcolor[rgb]{0.00,0.63,0.00}{##1}}}
\expandafter\def\csname PY@tok@gr\endcsname{\def\PY@tc##1{\textcolor[rgb]{1.00,0.00,0.00}{##1}}}
\expandafter\def\csname PY@tok@ge\endcsname{\let\PY@it=\textit}
\expandafter\def\csname PY@tok@gs\endcsname{\let\PY@bf=\textbf}
\expandafter\def\csname PY@tok@gp\endcsname{\let\PY@bf=\textbf\def\PY@tc##1{\textcolor[rgb]{0.00,0.00,0.50}{##1}}}
\expandafter\def\csname PY@tok@go\endcsname{\def\PY@tc##1{\textcolor[rgb]{0.53,0.53,0.53}{##1}}}
\expandafter\def\csname PY@tok@gt\endcsname{\def\PY@tc##1{\textcolor[rgb]{0.00,0.27,0.87}{##1}}}
\expandafter\def\csname PY@tok@err\endcsname{\def\PY@bc##1{\setlength{\fboxsep}{0pt}\fcolorbox[rgb]{1.00,0.00,0.00}{1,1,1}{\strut ##1}}}
\expandafter\def\csname PY@tok@kc\endcsname{\let\PY@bf=\textbf\def\PY@tc##1{\textcolor[rgb]{0.00,0.50,0.00}{##1}}}
\expandafter\def\csname PY@tok@kd\endcsname{\let\PY@bf=\textbf\def\PY@tc##1{\textcolor[rgb]{0.00,0.50,0.00}{##1}}}
\expandafter\def\csname PY@tok@kn\endcsname{\let\PY@bf=\textbf\def\PY@tc##1{\textcolor[rgb]{0.00,0.50,0.00}{##1}}}
\expandafter\def\csname PY@tok@kr\endcsname{\let\PY@bf=\textbf\def\PY@tc##1{\textcolor[rgb]{0.00,0.50,0.00}{##1}}}
\expandafter\def\csname PY@tok@bp\endcsname{\def\PY@tc##1{\textcolor[rgb]{0.00,0.50,0.00}{##1}}}
\expandafter\def\csname PY@tok@fm\endcsname{\def\PY@tc##1{\textcolor[rgb]{0.00,0.00,1.00}{##1}}}
\expandafter\def\csname PY@tok@vc\endcsname{\def\PY@tc##1{\textcolor[rgb]{0.10,0.09,0.49}{##1}}}
\expandafter\def\csname PY@tok@vg\endcsname{\def\PY@tc##1{\textcolor[rgb]{0.10,0.09,0.49}{##1}}}
\expandafter\def\csname PY@tok@vi\endcsname{\def\PY@tc##1{\textcolor[rgb]{0.10,0.09,0.49}{##1}}}
\expandafter\def\csname PY@tok@vm\endcsname{\def\PY@tc##1{\textcolor[rgb]{0.10,0.09,0.49}{##1}}}
\expandafter\def\csname PY@tok@sa\endcsname{\def\PY@tc##1{\textcolor[rgb]{0.73,0.13,0.13}{##1}}}
\expandafter\def\csname PY@tok@sb\endcsname{\def\PY@tc##1{\textcolor[rgb]{0.73,0.13,0.13}{##1}}}
\expandafter\def\csname PY@tok@sc\endcsname{\def\PY@tc##1{\textcolor[rgb]{0.73,0.13,0.13}{##1}}}
\expandafter\def\csname PY@tok@dl\endcsname{\def\PY@tc##1{\textcolor[rgb]{0.73,0.13,0.13}{##1}}}
\expandafter\def\csname PY@tok@s2\endcsname{\def\PY@tc##1{\textcolor[rgb]{0.73,0.13,0.13}{##1}}}
\expandafter\def\csname PY@tok@sh\endcsname{\def\PY@tc##1{\textcolor[rgb]{0.73,0.13,0.13}{##1}}}
\expandafter\def\csname PY@tok@s1\endcsname{\def\PY@tc##1{\textcolor[rgb]{0.73,0.13,0.13}{##1}}}
\expandafter\def\csname PY@tok@mb\endcsname{\def\PY@tc##1{\textcolor[rgb]{0.40,0.40,0.40}{##1}}}
\expandafter\def\csname PY@tok@mf\endcsname{\def\PY@tc##1{\textcolor[rgb]{0.40,0.40,0.40}{##1}}}
\expandafter\def\csname PY@tok@mh\endcsname{\def\PY@tc##1{\textcolor[rgb]{0.40,0.40,0.40}{##1}}}
\expandafter\def\csname PY@tok@mi\endcsname{\def\PY@tc##1{\textcolor[rgb]{0.40,0.40,0.40}{##1}}}
\expandafter\def\csname PY@tok@il\endcsname{\def\PY@tc##1{\textcolor[rgb]{0.40,0.40,0.40}{##1}}}
\expandafter\def\csname PY@tok@mo\endcsname{\def\PY@tc##1{\textcolor[rgb]{0.40,0.40,0.40}{##1}}}
\expandafter\def\csname PY@tok@ch\endcsname{\let\PY@it=\textit\def\PY@tc##1{\textcolor[rgb]{0.25,0.50,0.50}{##1}}}
\expandafter\def\csname PY@tok@cm\endcsname{\let\PY@it=\textit\def\PY@tc##1{\textcolor[rgb]{0.25,0.50,0.50}{##1}}}
\expandafter\def\csname PY@tok@cpf\endcsname{\let\PY@it=\textit\def\PY@tc##1{\textcolor[rgb]{0.25,0.50,0.50}{##1}}}
\expandafter\def\csname PY@tok@c1\endcsname{\let\PY@it=\textit\def\PY@tc##1{\textcolor[rgb]{0.25,0.50,0.50}{##1}}}
\expandafter\def\csname PY@tok@cs\endcsname{\let\PY@it=\textit\def\PY@tc##1{\textcolor[rgb]{0.25,0.50,0.50}{##1}}}

\def\PYZus{\char`\_}

\def\PYZgt{\char`\>}

\def\PYZdq{\char`\"}

\makeatother

    \definecolor{incolor}{rgb}{0.0, 0.0, 0.5}
    \definecolor{outcolor}{rgb}{0.545, 0.0, 0.0}

    \hypersetup{
      breaklinks=true,  
      colorlinks=true,
      urlcolor=urlcolor,
      linkcolor=linkcolor,
      citecolor=citecolor,
      }
    
\usepackage{thmtools}
\usepackage{relsize}
\usepackage{multirow}
\usepackage{multicol}
\usepackage{rotating}
\usepackage{stackengine}

\usetikzlibrary{arrows,
	automata,
	calc,
	decorations,
	decorations.pathreplacing,
	positioning,
	shapes}

\setenumerate[1]{label={(\arabic*)}} 
\numberwithin{equation}{section}
\npdecimalsign{.}
\npthousandsep{,}

\newcommand{\R}{\mathbb{R}}
\renewcommand{\P}{\mathcal{P}}
\newcommand{\prb}{\mathbf{Pr}}
\newcommand{\Ab}{\mathbf{A}}

\newcommand{\Pb}{\mathbf{P}}

\newcommand{\yb}{\mathbf{y}}
\newcommand{\bb}{\mathbf{b}}

\newcommand{\M}{\mathcal{M}}

\newcommand{\vb}{\mathbf{v}}

\newcommand{\zb}{\mathbf{z}}
\newcommand{\all}{{\operatorname{all}}}
\newcommand{\Act}{\operatorname{Act}}

\newcommand{\goal}{\operatorname{goal}}
\newcommand{\fail}{\operatorname{fail}}

\renewcommand{\S}{\mathfrak{S}}
\renewcommand{\Pr}{\mathrm{Pr}}

\newcommand{\dipro}{\textsc{Dipro}}
\newcommand{\comics}{\textsc{Comics}}
\newcommand{\ltlsubsys}{\texttt{ltlsubsys}}
\newcommand{\prism}{\textsc{Prism}}
\newcommand{\gurobi}{\textsc{Gurobi}}
\newcommand{\cbc}{\textsc{Cbc}}
\newcommand{\glpk}{\textsc{Glpk}}
\newcommand{\cplex}{\textsc{Cplex}}

\newcommand{\toolname}{{\scshape Switss}}
\newcommand{\code}[1]{\mbox{\small\texttt{#1}}}
\newcommand{\invF}{\operatorname{InvF}}
\newcommand{\invP}{\operatorname{InvP}}

\begin{document}
	
\title{\toolname: Computing Small\\ Witnessing Subsystems\\
	\thanks{This work was funded by DFG grant 389792660 as part of \href{https://perspicuous-computing.science}{TRR~248}, the Cluster of Excellence EXC 2050/1 (CeTI, project ID 390696704, as part of Germany's Excellence Strategy), DFG-projects BA-1679/11-1 and BA-1679/12-1, and the Research Training Group QuantLA (GRK 1763).}
}

\author{\IEEEauthorblockN{Simon Jantsch,
		Hans Harder, Florian Funke and
		Christel Baier}
	\IEEEauthorblockA{\\Technische Universität Dresden, Germany\\
		\{simon.jantsch, florian.funke, christel.baier\}@tu-dresden.de \\
		hans.harder@mailbox.tu-dresden.de}}
\maketitle	
	
\begin{abstract}
  Witnessing subsystems for probabilistic reachability thresholds in discrete Markovian models are an important concept both as diagnostic information on why a property holds, and as input to refinement algorithms.
  We present \toolname{}, a tool for the computation of \textbf{S}mall \textbf{WIT}nessing \textbf{S}ub\textbf{S}ystems.
  \toolname{} implements exact and heuristic approaches based on reducing the problem to (mixed integer) linear programming.
  Returned subsystems can automatically be rendered graphically and are accompanied with a certificate which proves that the subsystem is indeed a witness.
\end{abstract}

\section{Introduction}

A standard notion of a \emph{witness} for a property in probabilistic systems is that of a \emph{subsystem}  \cite{JansenAKWKB11,JansenWAZKBS14,BraitlingWBJA11,AbrahamBDJKW14,WimmerJAKB14,WimmerJABK12}. This is a part of the system that \emph{by itself} already reaches a given probability threshold and thus serves as an explanation of why or where the property holds. Subsystems can also be used as input to automated refinement and synthesis algorithms. In~\cite{HermannsWZ08} a \emph{counterexample guided abstraction refinement} (CEGAR) method for probabilistic models is presented that iteratively refines a predicate abstraction by analyzing counterexamples (which are witnessing subsystems to the negated property).
An application of small witnessing subsystems to synthesis is described in~\cite{CeskaHJK19}, where they are used to infer properties of a family of Markov chains from (a subsystem of) one of its members.

The aforementioned applications heavily benefit from witnessing subsystems that are \emph{small} in terms of their state space. This paper presents \toolname{}, a novel tool for the computation of small witnessing subsystems for reachability properties in Markovian models. Following \cite{FunkeJB20}, \toolname{} proceeds by reduction to finding points of a polyhedron containing a large number of zero entries. These points also serve as \emph{certificates} \cite{McConnellMNS11} for the fact that the computed subsystem indeed constitutes a witness.

We tackle the above problem from discrete geometry heuristically with an iterative linear programming (LP) approach. By adding binary variables to the LP, thus resulting in a mixed integer linear program (MILP), \toolname{} can also compute \emph{minimal} witnessing subsystems. In many applications, however, minimizing merely the state space of witnessing subsystems is insufficient in that it ignores the underlying structure of the model. For this reason, \toolname{} supports \emph{label-based minimization}, where syntactic units of the system can be subsumed under common labels.

Transparency and reliability are important factors for the evaluation of modern model checking software. \toolname{} comes with a toolkit for the automated visualization of Markovian models and subsystems therein. For the convenience of (third-party) users, the framework includes a separate module for the independent verification of the associated certificates. In this way, results can be checked both visually and mathematically.

The translation to discrete geometry paired with the high level of encapsulation in our implementation makes \toolname{} easily extendable. New heuristic approaches for finding vertices with many zeros (like vertex enumeration techniques) as well as different LP and MILP solvers used as backend engines can be integrated flexibly into \toolname{}.

\subsection*{Comparison with related tools.} 

There are, to the best of our knowledge, three existing tools for the computation of witnessing subsystems: \dipro{} \cite{AljazzarLLS11}, \comics{} \cite{JansenAVWKB12}, and \ltlsubsys{} \cite{WimmerJAKB14}. We now compare each of these to \toolname. A foreword that applies to all of them is that for Markov decision process (MDP), they only compute witnessing subsystems for \emph{lower} bounds on \emph{maximal} reachability probabilities. We emphasize that lower bounds on \emph{minimal} reachability probabilities cannot be reduced to this case, but only to \emph{upper} bounds on maximal reachability probabilities. 

 \dipro{} implements several heuristics for the computation of probabilistic counterexamples, only one of which directly operates on subsystems. This heuristic called XBF is available only for discrete-time Markov chains (DTMC), however. The other heuristics gather individual paths satisfying criteria like high probability mass or short length until the threshold is met. The subsystem resulting from these paths is not optimized along our state-minimality criterion (rather, the \emph{number of paths} is minimized), so a comparison is problematic. 
 
 \comics{} implements heuristics for computing small subsystems in DTMCs, which are significantly different from the ones implemented in \toolname{}.
 They rely on iteratively adding ``probable'' paths to the subsystem until the threshold is met.
 To compute the next such path, \comics{} uses graph algorithms.
 As each iteration requires computing the probability that has already been gathered, this approach often suffers from a substantial increase in time and memory consumption for growing thresholds, in contrast to our approach. With a prototypical implementation of what has now become \toolname{}, we found that either our minimal or maximal reachability formulation (both of which are available for DTMCs) usually outperforms both \comics{} modes \cite{FunkeJB20}.
 
\ltlsubsys{} \cite{WimmerJAKB14} is the only tool for the computation of \emph{minimal} witnesses in MDPs (it is, however, not publicly available).
This tool also reduces the task of computing witnessing subsystems for maximal probabilities to a MILP which is related to our MILP formulations (cf. \cite[Remark 6.2]{FunkeJB20}). Its results in terms of upper and lower bounds on the number of states in a minimal witness found when hitting the timeout (which usually happens for bigger models) are comparable to ours~\cite{FunkeJB20}. As mentioned above, \ltlsubsys{} cannot handle minimal reachability probabilities.

Summarizing the functionality, \toolname{} is the first tool that implements (1) both exact and heuristic algorithms, with support for (2) both DTMCs and, more generally, MDPs, for (3) thresholds on both minimal and maximal reachability probabilities.

\section{Theoretical background}\label{sec:background}
\renewcommand{\arraystretch}{1.5}
\renewcommand{\tabcolsep}{2mm}
\setlength{\textfloatsep}{0.3cm}

\begin{table}[tbp]
	\centering
	\caption{Overview of Farkas certificates for reachability properties in MDPs, where $\triangleleft \; \in \{\leq, <\}$ and $\triangleright \; \in\{\geq, > \}$.}\label{tab:farkcert}
  \begin{tabular}{ |c|c|c|}
		\hline
		Property& Source&  Condition  \\ \hline
		$\prb^{\min}_{s_0}(\lozenge {\text{\small goal}}) \triangleright \lambda$ & $\zb\in \R^{S}$ & $\Ab \zb \leq \bb \land \zb(s_0) \triangleright \lambda$  \\\hline
		$\prb^{\max}_{s_0}(\lozenge {\text{\small goal}}) \triangleright \lambda$ &  $\yb \in \R^{\M}_{\geq 0}$ & $\yb\Ab \leq\delta_{s_0}\land \yb\bb \triangleright \lambda$   \\\hline
		$\prb^{\min}_{s_0}(\lozenge {\text{\small goal}})\triangleleft \lambda$ & $\yb \in \R^{\M}_{\geq 0}$ & $\yb\Ab \geq\delta_{s_0}\land \yb\bb \triangleleft \lambda$   \\\hline
		$\prb^{\max}_{s_0}(\lozenge {\text{\small goal}}) \triangleleft \lambda$ & $\zb\in \R^{S}$ & $\Ab \zb \geq \bb \land \zb(s_0) \triangleleft \lambda$  \\\hline
	\end{tabular}
\end{table}

A \emph{Markov decision process} (MDP) is a tuple $\M = (S_\all, \Act, \Pb, s_0)$, where $S_\all$ is a finite set of \emph{states}, $\Act$ is a finite set of \emph{actions}, $\Pb\colon S_\all\times\Act\times S_\all\to [0,1]$ is the \emph{transition probability function} where we require $\sum_{s'\in S_\all} \Pb(s, \alpha, s') \in\{0,1\}$ for all $(s,\alpha)\in S_\all \times \Act$, and $s_0$ is the \emph{initial state} of $M$. We assume that there are two distinguished absorbing states $\fail, \goal\in S_\all$, representing desirable and undesirable outcomes of the system. We will henceforth use the notation $S = S_\all \setminus \{\fail, \goal\}$. We let $\Act(s)$ be the set of actions satisfying $\sum_{s'\in S_\all} \Pb(s, \alpha, s') =1$. We require $\Act(s) \neq \varnothing$ for all $s\in S$ and sometimes write $\M = \{(s,\alpha) \mid s \in S, \, \alpha \in \Act(s)\}$.

The system begins in $s_0$ and evolves as follows: in state $s$, an action $\alpha\in\Act(s)$ is chosen non-deterministically and the next state is picked according to the distribution $\Pb(s,\alpha, \cdot)$. A \emph{scheduler} $\S$ is some resolution of the non-determinism and induces a probability $\Pr^{\S}_{\M, s_0}(\lozenge\goal)$ to eventually reach $\goal$ (see \cite[Section 10.6]{BaierK2008}). 
We are interested in the \emph{minimal} and \emph{maximal} reachability probabilities attained among all schedulers, denoted by $\prb^{\min}_{\M, s_0}(\lozenge\goal)$ and $\prb^{\max}_{\M, s_0}(\lozenge\goal)$. They represent worst- and best-case scenarios for the behavior of the system.

A \emph{subsystem} of $\M$ is an MDP obtained from $\M$ by deleting states from $S$ and redirecting transitions to $\fail$. If $\M$ satisfies $\prb^{*}_{\M, s_0}(\lozenge\goal)\geq\lambda$ for $*\in\{\min, \max\}$, one way of analyzing which parts of the system are sufficient for this inequality is to find a subsystem $\M'$ of $\M$ already satisfying the lower bound, i.e., $\prb^{*}_{\M', s_0}(\lozenge\goal)\geq\lambda$. We call these \emph{witnessing subsystems}. We aim at finding small (or \emph{minimal}) witnessing subsystems in terms of how many states they include.

In \cite{FunkeJB20} we proposed a translation between witnessing subsystems and \emph{Farkas certificates} (which are vectors satisfying the conditions in~\Cref{tab:farkcert}) for lower-bounded reachability thresholds. Here, $\mathbf{A} \in \mathbb{R}^{\M \times S}$ and $\bb \in \R^S$ are defined as follows: $\mathbf{A}((s,\alpha),t) = 1 - \Pb(s,\alpha,s)$ if $s = t$ and $-\Pb(s,\alpha,t)$ otherwise,  and $\bb(s, \alpha) = \Pb(s, \alpha, \goal)$.

  In this paper, we are mainly interested in the first two rows of \Cref{tab:farkcert} with $\triangleright = \, \geq$, and denote the corresponding sets of Farkas certificates by $\P^{\min}_{\M}(\lambda) \subseteq \R^{S}_{\geq 0}$ and $ \P^{\max}_{\M}(\lambda) \subseteq \R^{\M}_{\geq 0}$. The passage from a Farkas certificate $\zb \in \P^{\min}_{\M}(\lambda)$ (resp. $\yb \in \P^{\max}_{\M}(\lambda)$) to a subsystem of $\M$ works by including all states with $\zb(s)>0$ (resp. $\yb(s, \alpha)>0$ for some $\alpha$), and all edges between such states. All other edges are redirected to $\fail$. Thus, computing minimal (small) witnessing subsystems for $\prb^{*}_{\M, s_0}(\lozenge\goal)\geq\lambda$ can be reduced to finding points in $\P^{*}_{\M}(\lambda)$ with a maximal (large) number of zeros.

As in~\cite{FunkeJB20} we have to assume that the only \emph{maximal end components} of $\M$ are $\{\goal\}$ and $\{\fail\}$. This means that almost all paths reach either of these two states under every scheduler.
This can be ensured by a preprocessing step whose time-complexity is at most quadratic in the underlying graph, see~\cite{ChatterjeeH11, deAlfaro97b}.

\section{Implementation and features}

\begin{figure*}[tbp]
  \centering
    \scalebox{1.25}{\includegraphics{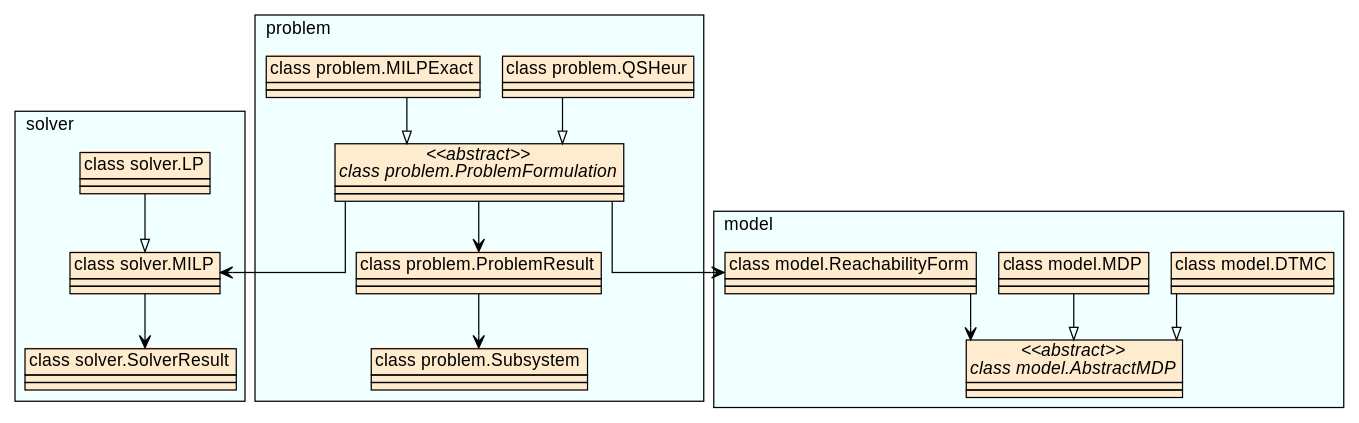}}
  \caption{\toolname{} contains modules for modeling DTMCs and MDPs (\code{switss.model}), different approaches to finding small (or minimal) subsystems (\code{switss.problem}) and interfaces to MILP and LP solvers, built on top of the PuLP library (\code{switss.solver}). The \code{model.ReachabilityForm} class is a wrapper for DTMCs/MDPs which fulfill the requirements as given in \Cref{sec:background} (having a single target and fail state, etc.). Finding small subsystems is done through implementations of \code{problem.ProblemFormulation}, e.g. by using the quotient sum heuristics (\code{problem.QSHeur}) or the MILP formulation (\code{problem.MILPExact}). Additionally included are modules for benchmarking (\code{switss.benchmarks}), generating and verifying Farkas certificates (\code{switss.certfication}) and interaction with \prism{} and \prism{} file formats (\code{switss.prism}).}
  \label{fig:uml}
\end{figure*}

\toolname{}\footnote{\url{https://github.com/simonjantsch/switss}} is a complete re-implementation and substantial extension of the prototype implementation that was used to run the experiments presented in~\cite{FunkeJB20}.
An overview of the structure of \toolname{} is given in~\Cref{fig:uml}.
Apart from increased usability and an extensive documentation and testing suite, the main extensions are the following:
\begin{itemize}
\item Functions to generate and verify certificates for all senses ($\leq,<,\geq,>$) and modes (min/max).
\item Visualization of MDP subsystems.
\item A generalized notion of minimality that allows to minimize ``active labels'' \cite{WimmerJAK15} in the subsystem (both exactly and heuristically).
\item New heuristics for the computation of witnessing subsystems with few states.
\item Support for various LP/MILP solvers.
\end{itemize}

\toolname{} is implemented in python and can be included as a library, or used as a stand-alone tool.
MDPs and DTMCs can be loaded either from the explicit transition matrix format (.tra), or from a model specified in the \prism~\cite{KwiatkowskaNP11} guarded command language.
In the latter case, \toolname{} uses \prism{} to derive an explicit transition matrix representation.
The library PuLP\footnote{\url{https://coin-or.github.io/pulp/}} is used as modeling language for linear programs, and as an interface to various LP/MILP solvers, where we currently support: \gurobi~\cite{gurobi}, \cplex\footnote{\url{https://www.ibm.com/analytics/cplex-optimizer}}, \cbc\footnote{\url{https://github.com/coin-or/Cbc}} and \glpk\footnote{\url{https://www.gnu.org/software/glpk/}}.
While the first two are proprietary software (both offer academic licenses), the latter two are open source.
 
\subsection{Computing and verifying Farkas certificates}
Generating Farkas certificates (for a specified threshold, sense and mode) amounts to finding a vector satisfying the corresponding linear inequalities as presented in~\Cref{tab:farkcert}.
For the non-strict inequalities, this can be done directly by solving an LP, where the objective function can be arbitrary. 
Handling strict inequalities can be done by replacing the strict inequality by its non-strict counterpart, and then optimizing in the direction where strict inequality is required.
A Farkas certificate exists if and only if the solution satisfies the strict inequality (and the solution is then a certificate).

To verify that a given vector $\vb$ is a Farkas certificate, it is enough to check that it satisfies the inequalities.
Due to the varying precision of solvers and the python numerical libraries, it can happen that exact satisfaction of the certificate condition is not given.
Hence we allow a \emph{tolerance} $t$ to be passed as an option to the certificate verifier, which will then check, for example, that:
\[\Ab \vb - \mathbf{t} \leq \bb \land \vb(s_0) + t \geq \lambda\]
where $\mathbf{t}$ is the vector of appropriate dimension containing $t$ in every entry.
In the future, we plan to explore how \emph{robust} certificates (which can be verified with $t=0$) can be generated efficiently and consistently (e.g. by searching for vectors that do not lie on the boundary of the polytope).

\subsection{New heuristics}

To find points in $\P^*_{\M}(\lambda)$ with many zeros, the \emph{quotient-sum} (QS) heuristic~\cite{FunkeJB20} iteratively solves LPs over the polytope $\P^*_{\M}(\lambda)$.
The objective function is updated in every step in a way that aims at pushing as many entries of the solutions vectors to zero as possible.
The LP that is solved for $\P_{\M}^{\min}(\lambda)$ is:
\begin{equation}
  \min \sum_{s \in S}\mathbf{o}_i(s) \zb(s) \;\; \text{ s.t. } \quad \zb \in \P^{\min}_{\M}(\lambda)
\end{equation}
We put $\mathbf{o}_0 = (1,\ldots,1) \in \mathbb{R}^S$ and compute $\mathbf{o}_{j+1}$ from a solution $\zb_j$ of the $j$-th iteration by the quotient rule: $\mathbf{o}_{j+1}(s) = 1/\zb_j(s)$ if $\zb_j(s) > 0$, and else $\mathbf{o}_{j+1}(s) = C$ for some big $C$.
Hence, a dimension close to zero will have a high ``cost'' in the next iteration.

While the heuristics generally yield small subsystems fast (especially when compared with the time it takes to exactly minimize the number of states with a MILP), sometimes ``spikes'' were observed: the heuristics returned a worse result when decreasing $\lambda$.
As a witnessing subsystem for $\lambda$ is also witnessing for all $\lambda' \leq \lambda$, this is undesirable.

In \toolname{} both the initial objective $\mathbf{o}_0$ and the update can be customized.
As shown by our experiments (\Cref{sec:experiments}), the choice of $\mathbf{o}_0$ may have a substantial effect on the performance of the heuristics (first experiments on changing the update did not lead to better performance).
We propose the following candidate values for $\mathbf{o}_0$.
For the heuristic related to $\prb^{\max}$, we take $\mathbf{o}_0$ to be the inverse of a solution to the following LP (where the inverse is the result of pointwise $1/\cdot$, if the corresponding entry is greater 0, and a big constant otherwise):
\begin{equation}
  \label{eqn:maxy}
  \max \, \yb \bb \;\; \text{\footnotesize s.t. } \;\; \yb \in \P^{\max}_{\M}(0)
\end{equation}
For $\prb^{\min}$ we let $\mathbf{o}_0$ be the inverse of a solution to:
\begin{equation}
  \label{eqn:maxz}
  \max \sum_{s \in S} \zb(s) \;\; \text{\footnotesize s.t. } \;\; \zb \in \P^{\min}_{\M}(0)
\end{equation}
Putting $\lambda =0$ discards the constraint $\zb(s_0) \geq \lambda$ for $* = \min$ and $\yb \bb \geq \lambda$ for $* = \max$ (compare~\Cref{tab:farkcert}).
If $\M$ is a DTMC, then a solution vector of \Cref{eqn:maxy} contains the \emph{expected number of visits to a state} and a solution of \Cref{eqn:maxz} contains the probability to reach $\goal$ from every state.
Intuitively, states with a low entry in these vectors contribute less to the probability of reaching $\goal$ from the initial state, and hence should get a higher value in $\mathbf{o}_0$.
Similar importance measures where considered in the context of counterexample generation in~\cite{BrazdilCCFK2015}.

\subsection{Label-based minimization}
The idea of minimizing not the number of states, but the number of labels present in a subsystem was first considered in~\cite{WimmerJAK15}.
There it was used to minimize the number of ``active'' commands for MDPs given in \prism{} language.
We have extended the approach of~\cite{FunkeJB20} to allow label-based minimization in a similar way as was done in~\cite{WimmerJAK15}.
This allows applying the QS-heuristic to the computation of subsystems with few active labels.
Further interesting use cases could be minimizing participating components (for compositional systems) in a witnessing subsystem, or the number of controllable states.

Our extension works as follows. Take an MDP $\M$ with states $S \cup \{\goal,\fail\}$ a finite set of labels $L$ and $\Lambda : S \to 2^L$.
Now let $\sigma$ be a vector with $|L|$ variables with domain $[0,1]$ and consider the following LPs, which generalize the LPs of~\cite[Section 6]{FunkeJB20}:
\begin{equation}
  \label{eqn:minlabels}
  \min \sum_{l \in L} \sigma(l) \text{ \footnotesize s.t.} \;\;\;  \stackanchor{$\zb \in \P^{\min}_{\M}(\lambda)\;\;$ }{$\zb(s) \leq \sigma(l) \,\;\;\; \text{ \footnotesize f.a. } \;\, {\substack{s\in S \\ l \in \Lambda(s)}}$}
\end{equation}
\begin{equation}
  \label{eqn:maxlabels}
  \min \sum_{l \in L} \sigma(l) \text{ \footnotesize s.t.} \; \stackanchor{$\yb \in \P^{\max}_{\M}(\lambda) \;\,\;\;\;\;\,$ }{$\yb(s,\alpha) \leq K \cdot \sigma(l) \, \text{ \footnotesize f.a. } {\substack{(s,\alpha) \in \M \\ l \in \Lambda(s)}}$}
\end{equation}
The factor $K$ in~\Cref{eqn:maxlabels} is an upper bound on any entry of any vector $\yb \in \P_\M^{\max}(\lambda)$ (here we use that $\P_\M^{*}(\lambda)$ is bounded, cf. \cite[Lemma 5.1]{FunkeJB20}). 
It can be computed by first maximizing the sum of all entries over all vectors in $\P_\M^{\max}(\lambda)$ using an LP, and taking the objective value of the solution to be $K$.
The first LP does not need this step, as $1$ is an upper bound on all entries in any vector $\zb \in \P_{\M}^{\min}(\lambda)$.

A solution $(\zb,\sigma)$ of \Cref{eqn:minlabels} with $N$ non-zero entries in $\sigma$ can be translated into a witnessing subsystem for $\prb^{\min}_{\M}(\lozenge \goal) \geq \lambda$ with $N$ labels.
Conversely, a witnessing subsystem with $N$ active labels induces a solution $(\zb,\sigma)$ such that $\sigma$ has $N$ non-zero entries.
The same holds for \Cref{eqn:maxlabels}.

The QS-heuristic can be adapted to label-based minimization by trying to push only the entries of $\sigma$ to zero. This algorithm is implemented in \toolname{}.
Restricting the domain of $\sigma$-variables in \Cref{eqn:minlabels,eqn:maxlabels} to $\{0,1\}$ yields a MILP, whose solutions correspond to witnessing subsystems with a minimal amount of present labels.

\begin{figure}[tbp]
  \centering
    \resizebox{!}{9.5cm}{
      \includegraphics{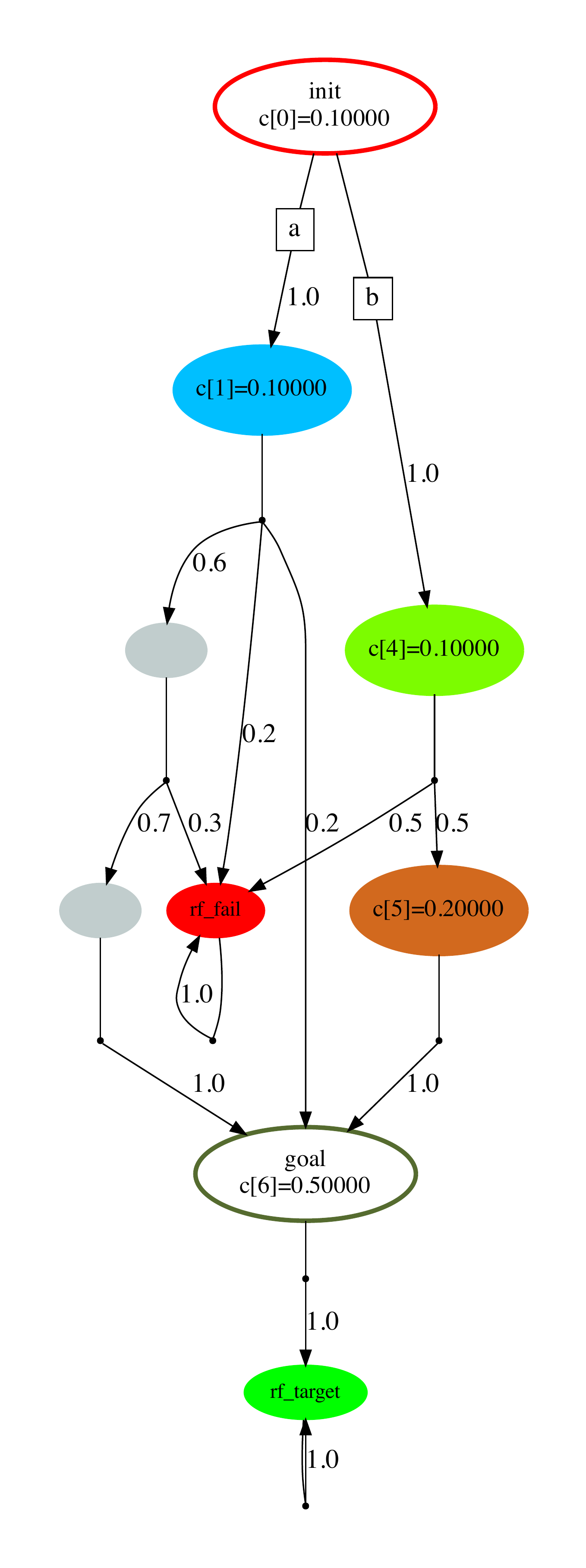}}
  \caption{Visualization of a subsystem (which excludes the gray states) by \toolname{} .
    The other colors indicate a user defined labeling.
    If the subsystem is induced by a Farkas certificate (e.g. as returned by the QS-heuristic) \toolname{} prints the corresponding values in each state (or action, for ``max''-queries). }
  \label{fig:outputexample}
\end{figure}

\section{A tour of \toolname}
The following tour can be reproduced using the supplementary material~\cite{FMCAD20-supplementary}. 
We first load an MDP:

    \begin{Verbatim}[commandchars=\\\{\},fontsize=\fontsize{8}{9}]
{\color{incolor}In:} \PY{n}{mc} \PY{o}{=} \PY{n}{MDP}\PY{o}{.}\PY{n}{from\PYZus{}file}\PY{p}{(}
        \PY{l+s+s2}{\PYZdq{}}\PY{l+s+s2}{ex\PYZus{}mdp.lab}\PY{l+s+s2}{\PYZdq{}}\PY{p}{,} \PY{l+s+s2}{\PYZdq{}}\PY{l+s+s2}{ex\PYZus{}mdp.tra}\PY{l+s+s2}{\PYZdq{}}\PY{p}{)}
\end{Verbatim}
    The following command renders the MDP (we have rebuilt the MDP in tikz for a better presentation and refer to \Cref{fig:outputexample} for an example output of \toolname{}):
    \begin{Verbatim}[commandchars=\\\{\},fontsize=\fontsize{8}{9}]
{\color{incolor}In:} \PY{n}{mc}\PY{o}{.}\PY{n}{digraph}\PY{p}{(}\PY{p}{)}
\end{Verbatim}

\noindent{\fontsize{8}{9}{\texttt{\color{outcolor}Out:}}}
    
    \begin{center}
    \adjustimage{max size={0.9\linewidth}{0.12\paperheight}}{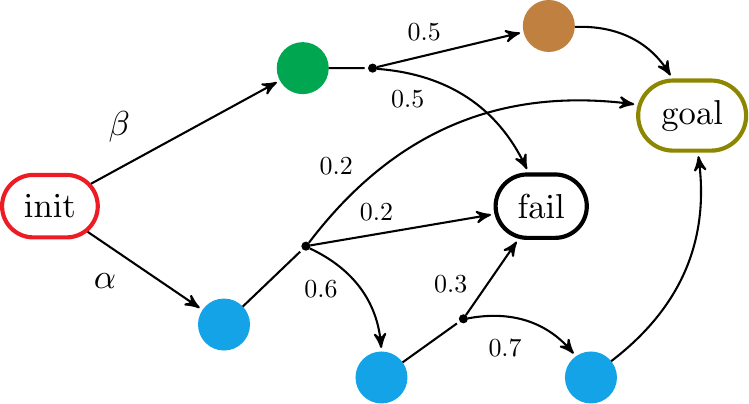}
    \end{center}
We first transform the MDP into so-called \emph{reachability form} (RF), which can be thought of as a standardized format for reachability analysis. It can be constructed from a DTMC or MDP, an initial state (which should be unique), and a set of target states.
The method \code{reduce} performs forward (from the intial
state) and backward (from the target states) reachability queries and
removes all states that are either unreachable, or do not reach $\goal$. A new distinguished target
state is added, which receives an incoming edge from each original target state.

    \begin{Verbatim}[commandchars=\\\{\},fontsize=\fontsize{8}{9}]
{\color{incolor}In:} \PY{n}{rf}\PY{p}{,}\PY{n}{\PYZus{}}\PY{p}{,}\PY{n}{\PYZus{}} \PY{o}{=} \PY{n}{ReachabilityForm}\PY{o}{.}\PY{n}{reduce}\PY{p}{(}
        \PY{n}{mc}\PY{p}{, }\PY{l+s+s2}{\PYZdq{}}\PY{l+s+s2}{init}\PY{l+s+s2}{\PYZdq{}}\PY{p}{, }\PY{l+s+s2}{\PYZdq{}}\PY{l+s+s2}{goal}\PY{l+s+s2}{\PYZdq{}}\PY{p}{)}
    \PY{n}{rf}\PY{o}{.}\PY{n}{system}\PY{o}{.}\PY{n}{digraph}\PY{p}{(}\PY{p}{)}
\end{Verbatim}

\noindent{\fontsize{8}{9}{\texttt{\color{outcolor}Out:}}}
    
    \begin{center}
    \adjustimage{max size={0.9\linewidth}{0.3\paperheight}}{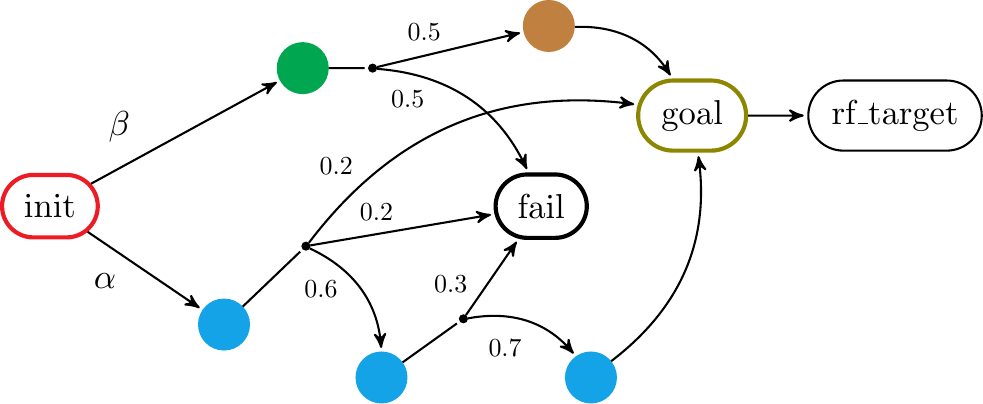}
    \end{center}

    \hypertarget{certification}{%
\subsection{Certification}\label{certification}}

    Next we demonstrate how Farkas certificates can be generated (by \code{generate\_farkas\_certificate}) and verified (by \code{check\_farkas\_certificate}). These methods take an RF and a
specification of the threshold property to be certified. We first generate a certificate using \cbc{} and verify its validity.
    \begin{Verbatim}[commandchars=\\\{\},fontsize=\fontsize{8}{9}]
{\color{incolor}In:} \PY{n}{cert} \PY{o}{=} \PY{n}{generate\PYZus{}farkas\PYZus{}certificate}\PY{p}{(}
        \PY{n}{rf}\PY{p}{, }\PY{l+s+s2}{\PYZdq{}}\PY{l+s+s2}{max}\PY{l+s+s2}{\PYZdq{}}\PY{p}{, }\PY{l+s+s2}{\PYZdq{}}\PY{l+s+s2}{\PYZgt{}=}\PY{l+s+s2}{\PYZdq{}}\PY{p}{, }\PY{l+m+mf}{0.55}\PY{p}{, }\PY{n}{solver}\PY{o}{=}\PY{l+s+s2}{\PYZdq{}}\PY{l+s+s2}{cbc}\PY{l+s+s2}{\PYZdq{}}\PY{p}{)}
\end{Verbatim}

    \begin{Verbatim}[commandchars=\\\{\},fontsize=\fontsize{8}{9}]
{\color{incolor}In:} \PY{n}{check\PYZus{}farkas\PYZus{}certificate}\PY{p}{(}
        \PY{n}{rf}\PY{p}{, }\PY{l+s+s2}{\PYZdq{}}\PY{l+s+s2}{max}\PY{l+s+s2}{\PYZdq{}}\PY{p}{, }\PY{l+s+s2}{\PYZdq{}}\PY{l+s+s2}{\PYZgt{}=}\PY{l+s+s2}{\PYZdq{}}\PY{p}{, }\PY{l+m+mf}{0.55}\PY{p}{, }\PY{n}{cert}\PY{p}{)}
\end{Verbatim}

\begin{Verbatim}[commandchars=\\\{\},fontsize=\fontsize{8}{9}]
{\color{outcolor}Out:} True
\end{Verbatim}

    If the threshold property is not satisfied by the model, no Farkas
certificate can be produced.

    \begin{Verbatim}[commandchars=\\\{\},fontsize=\fontsize{8}{9}]
{\color{incolor}In:} \PY{n}{fark\PYZus{}cert} \PY{o}{=} \PY{n}{generate\PYZus{}farkas\PYZus{}certificate}\PY{p}{(}
        \PY{n}{rf}\PY{p}{, }\PY{l+s+s2}{\PYZdq{}}\PY{l+s+s2}{min}\PY{l+s+s2}{\PYZdq{}}\PY{p}{, }\PY{l+s+s2}{\PYZdq{}}\PY{l+s+s2}{\PYZgt{}=}\PY{l+s+s2}{\PYZdq{}}\PY{p}{, }\PY{l+m+mf}{0.55}\PY{p}{, }\PY{n}{solver}\PY{o}{=}\PY{l+s+s2}{\PYZdq{}}\PY{l+s+s2}{cbc}\PY{l+s+s2}{\PYZdq{}}\PY{p}{)}
\end{Verbatim}

    \begin{Verbatim}[commandchars=\\\{\},fontsize=\fontsize{8}{9}]
{\color{outcolor}Out:} Property is not satisfied!
    \end{Verbatim}

    \hypertarget{witnessing-subsystems}{%
\subsection{Witnessing subsystems}\label{witnessing-subsystems}}

    We illustrate the computation of witnessing subsystems, beginning with the methods for exactly minimizing the number of states. In contrast to the certification module, we
only consider lower-bounded thresholds here, as reachability probabilities cannot increase in subsystems.

    \hypertarget{minimal-ws}{%
\subsubsection{Minimal witnessing subsystems}\label{minimal-ws}}
    The class \code{MILPExact} (an instance of \code{ProblemFormulation}) is used to specify queries for exact
minimization of an RF for a given threshold property. It is initialized
by specifying \emph{mode} (min or max) and solver:
\begin{Verbatim}[commandchars=\\\{\},fontsize=\fontsize{8}{9}]
{\color{incolor}In:} \PY{n}{milp\PYZus{}exact\PYZus{}max} \PY{o}{=} \PY{n}{MILPExact}\PY{p}{(}
        \PY{l+s+s2}{\PYZdq{}}\PY{l+s+s2}{max}\PY{l+s+s2}{\PYZdq{}}\PY{p}{, }\PY{n}{solver}\PY{o}{=}\PY{l+s+s2}{\PYZdq{}}\PY{l+s+s2}{cbc}\PY{l+s+s2}{\PYZdq{}}\PY{p}{)}
\end{Verbatim}

    The \code{solve} method now takes an RF and a threshold, constructs the
MILP and solves it by calling the specified solver.

    \begin{Verbatim}[commandchars=\\\{\},fontsize=\fontsize{8}{9}]
{\color{incolor}In:} \PY{n}{res} \PY{o}{=} \PY{n}{milp\PYZus{}exact\PYZus{}max}\PY{o}{.}\PY{n}{solve}\PY{p}{(}\PY{n}{rf}\PY{p}{, }\PY{l+m+mf}{0.1}\PY{p}{)}
\end{Verbatim}

    If successful, the result contains a subsystem that can also be rendered graphically, where pale states do not belong to the subsystem:

    \begin{Verbatim}[commandchars=\\\{\},fontsize=\fontsize{8}{9}]
{\color{incolor}In:} \PY{n}{res}\PY{o}{.}\PY{n}{subsystem}\PY{o}{.}\PY{n}{digraph}\PY{p}{(}\PY{p}{)}
\end{Verbatim}

\noindent{\fontsize{8}{9}{\texttt{\color{outcolor}Out:}}}
    
    \begin{center}
    \adjustimage{max size={0.9\linewidth}{0.3\paperheight}}{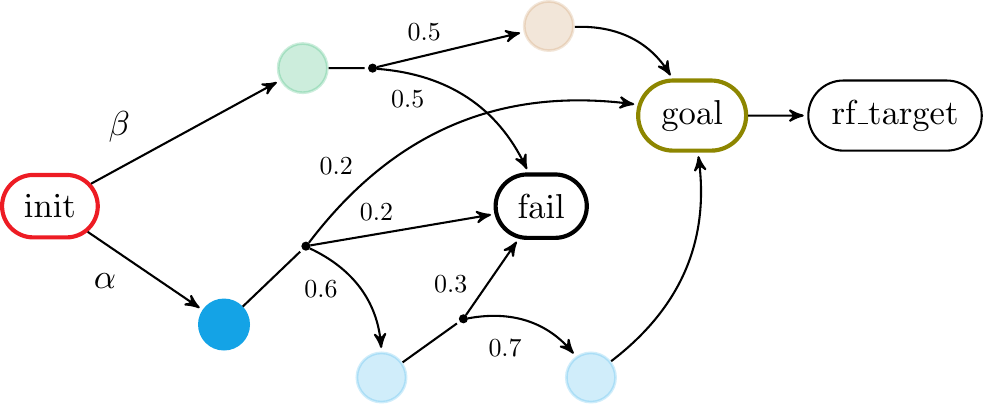}
    \end{center}

   If the threshold is increased to $0.3$, the minimal witnessing subsystem uses the upper branch:

    \begin{Verbatim}[commandchars=\\\{\},fontsize=\fontsize{8}{9}]
{\color{incolor}In:} \PY{n}{res} \PY{o}{=} \PY{n}{milp\PYZus{}exact\PYZus{}max}\PY{o}{.}\PY{n}{solve}\PY{p}{(}\PY{n}{rf}\PY{p}{, }\PY{l+m+mf}{0.3}\PY{p}{)}
\end{Verbatim}

    \begin{Verbatim}[commandchars=\\\{\},fontsize=\fontsize{8}{9}]
{\color{incolor}In:} \PY{n}{res}\PY{o}{.}\PY{n}{subsystem}\PY{o}{.}\PY{n}{digraph}\PY{p}{(}\PY{p}{)}
\end{Verbatim}

\noindent{\fontsize{8}{9}{\texttt{\color{outcolor}Out:}}}
    
    \begin{center}
    \adjustimage{max size={0.9\linewidth}{0.3\paperheight}}{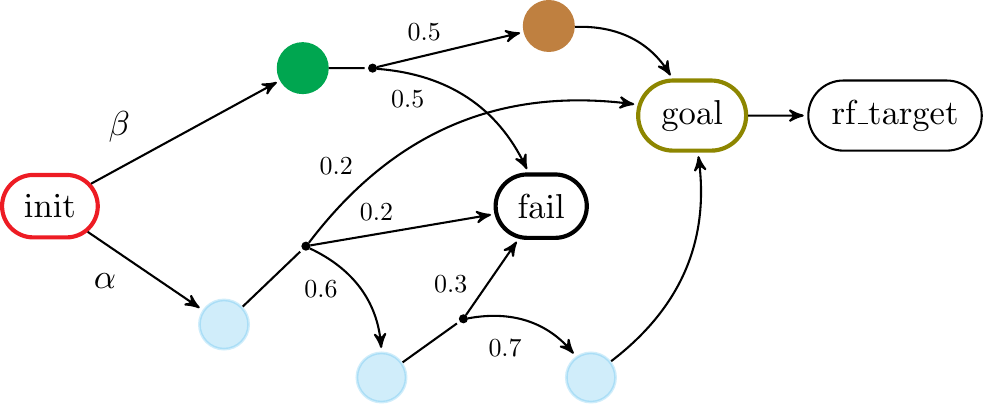}
    \end{center}




    

    We now consider witnesses for minimal reachability. Such a witness needs
to ensure that the threshold is met by \emph{all} possible schedulers.
For \(0.1\), it is enough to include the upper branch and the first state in the lower
branch, but for \(0.3\) already all states have to be included.

    \begin{Verbatim}[commandchars=\\\{\},fontsize=\fontsize{8}{9}]
{\color{incolor}In:} \PY{n}{milp\PYZus{}exact\PYZus{}min} \PY{o}{=} \PY{n}{MILPExact}\PY{p}{(}
        \PY{l+s+s2}{\PYZdq{}}\PY{l+s+s2}{min}\PY{l+s+s2}{\PYZdq{}}\PY{p}{, }\PY{n}{solver}\PY{o}{=}\PY{l+s+s2}{\PYZdq{}}\PY{l+s+s2}{cbc}\PY{l+s+s2}{\PYZdq{}}\PY{p}{)}
\end{Verbatim}

    \begin{Verbatim}[commandchars=\\\{\},fontsize=\fontsize{8}{9}]
{\color{incolor}In:} \PY{n}{res} \PY{o}{=} \PY{n}{milp\PYZus{}exact\PYZus{}min}\PY{o}{.}\PY{n}{solve}\PY{p}{(}\PY{n}{rf}\PY{p}{, }\PY{l+m+mf}{0.1}\PY{p}{)}
    \PY{n}{res}\PY{o}{.}\PY{n}{subsystem}\PY{o}{.}\PY{n}{digraph}\PY{p}{(}\PY{p}{)}
\end{Verbatim}

\noindent{\fontsize{8}{9}{\texttt{\color{outcolor}Out:}}}
    
    \begin{center}
    \adjustimage{max size={0.9\linewidth}{0.3\paperheight}}{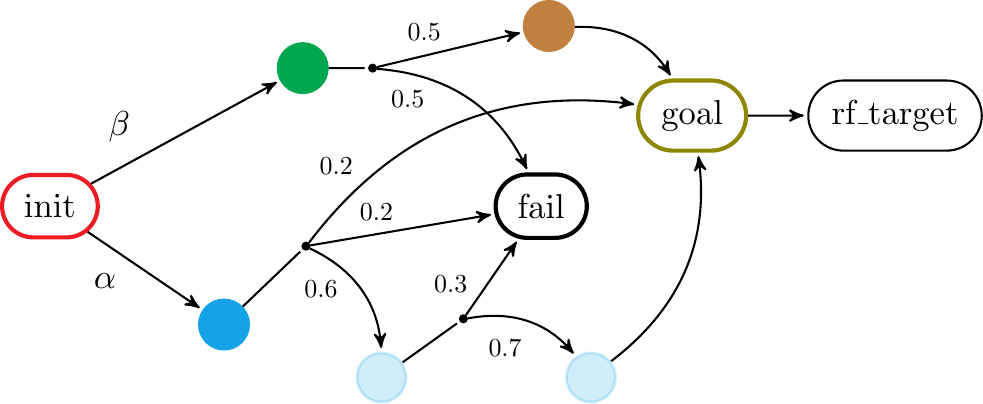}
    \end{center}

    \hypertarget{heuristics}{%
\subsubsection{The QS-heuristic}\label{heuristics}}

    The counterpart of \code{MILPExact} for the heuristic computations of small (rather than minimal) witnessing subsystems is
\code{QSHeur}. As it is an iterative heuristic, we can use the method
\code{solveiter} to return an iterator over the results. In our example,
all three iterations return the same subsystem, which is not
optimal, however (compare with the results of the exact query for maximum
probability and threshold \(0.1\)).

\begin{Verbatim}[commandchars=\\\{\},fontsize=\fontsize{8}{9}]
{\color{incolor}In:} \PY{n}{qs\PYZus{}max\PYZus{}heur} \PY{o}{=} \PY{n}{QSHeur}\PY{p}{(}
        \PY{l+s+s2}{\PYZdq{}}\PY{l+s+s2}{max}\PY{l+s+s2}{\PYZdq{}}\PY{p}{, }\PY{n}{solver}\PY{o}{=}\PY{l+s+s2}{\PYZdq{}}\PY{l+s+s2}{cbc}\PY{l+s+s2}{\PYZdq{}}\PY{p}{, }\PY{n}{iterations}\PY{o}{=}\PY{l+m+mi}{3}\PY{p}{)}
    \PY{n}{results} \PY{o}{=} \PY{n+nb}{list}\PY{p}{(}
        \PY{n}{qs\PYZus{}max\PYZus{}heur}\PY{o}{.}\PY{n}{solveiter}\PY{p}{(}\PY{n}{rf}\PY{p}{, }\PY{l+m+mf}{0.1}\PY{p}{)}\PY{p}{)}
    \PY{n}{print\PYZus{}results}\PY{p}{(}\PY{n}{results}\PY{p}{)}
\end{Verbatim}

    \begin{Verbatim}[commandchars=\\\{\}, fontsize=\fontsize{8}{9}]
{\color{outcolor}Out:} -- results --
subsys states:5, value: 5
subsys states:5, value: 5
subsys states:5, value: 5
    \end{Verbatim}

    \begin{Verbatim}[commandchars=\\\{\},fontsize=\fontsize{8}{9}]
{\color{incolor}In:} \PY{n}{results}\PY{p}{[}\PY{l+m+mi}{2}\PY{p}{]}\PY{o}{.}\PY{n}{subsystem}\PY{o}{.}\PY{n}{digraph}\PY{p}{(}\PY{p}{)}
\end{Verbatim}

\noindent{\fontsize{8}{9}{\texttt{\color{outcolor}Out:}}}
    
    \begin{center}
    \adjustimage{max size={0.9\linewidth}{0.3\paperheight}}{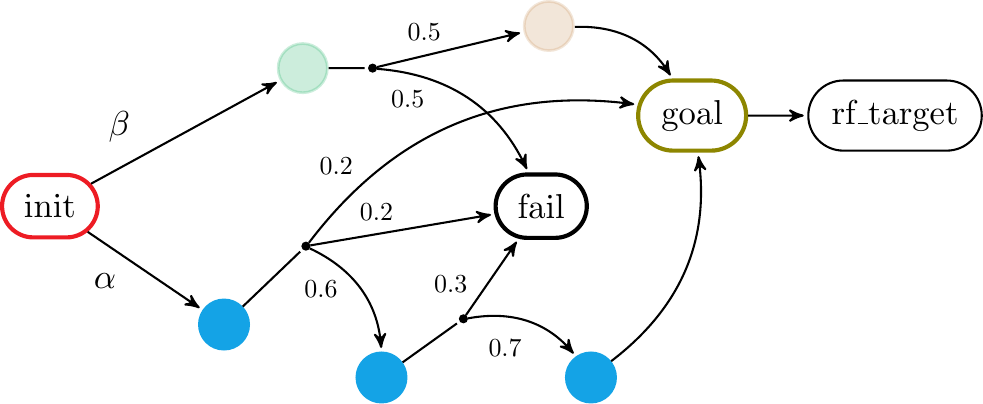}
    \end{center}

    The corresponding computation for minimum probabilities improves after
the first iteration and returns the optimal witness (the default number of iterations of the heuristics is three).

\begin{Verbatim}[commandchars=\\\{\},fontsize=\fontsize{8}{9}]
{\color{incolor}In:} \PY{n}{qs\PYZus{}min\PYZus{}heur} \PY{o}{=} \PY{n}{QSHeur}\PY{p}{(}
        \PY{l+s+s2}{\PYZdq{}}\PY{l+s+s2}{min}\PY{l+s+s2}{\PYZdq{}}\PY{p}{, }\PY{n}{solver}\PY{o}{=}\PY{l+s+s2}{\PYZdq{}}\PY{l+s+s2}{cbc}\PY{l+s+s2}{\PYZdq{}}\PY{p}{)}
    \PY{n}{results} \PY{o}{=} \PY{n+nb}{list}\PY{p}{(}
        \PY{n}{qs\PYZus{}min\PYZus{}heur}\PY{o}{.}\PY{n}{solveiter}\PY{p}{(}\PY{n}{rf}\PY{p}{, }\PY{l+m+mf}{0.1}\PY{p}{)}\PY{p}{)}
    \PY{n}{print\PYZus{}results}\PY{p}{(}\PY{n}{results}\PY{p}{)}
\end{Verbatim}

    \begin{Verbatim}[commandchars=\\\{\},fontsize=\fontsize{8}{9}]
{\color{outcolor}Out:} -- results --
subsys states:7, value: 7
subsys states:5, value: 5
subsys states:5, value: 5

    \end{Verbatim}

    \begin{Verbatim}[commandchars=\\\{\},fontsize=\fontsize{8}{9}]
{\color{incolor}In:} \PY{n}{results}\PY{p}{[}\PY{l+m+mi}{2}\PY{p}{]}\PY{o}{.}\PY{n}{subsystem}\PY{o}{.}\PY{n}{digraph}\PY{p}{(}\PY{p}{)}
\end{Verbatim}
\vspace{3mm}
\noindent{\fontsize{8}{9}{\texttt{\color{outcolor}Out:}}}
    
    \begin{center}
    \adjustimage{max size={0.9\linewidth}{0.3\paperheight}}{running_example/mdp_sub_145.pdf}
    \end{center}

    \hypertarget{taking-labels-into-account}{%
\subsection{Label-based minimization}\label{taking-labels-into-account}}

    Now suppose that we do not want to minimize the amount of states present
in the subsystem, but the amount of \emph{colors} that it includes.
The colors stand for some labeling that may interest the user.
That is, the lower branch counts as one, as it only includes one color,
while the upper branch counts two although it has a state less. We
specify this optimization objective by using the \code{labels}
parameter of the \code{solve} method.

        \begin{Verbatim}[commandchars=\\\{\},fontsize=\fontsize{8}{9}]
{\color{incolor}In:} \PY{n}{milp\PYZus{}exact\PYZus{}labels} \PY{o}{=} \PY{n}{MILPExact}\PY{p}{(}
        \PY{l+s+s2}{\PYZdq{}}\PY{l+s+s2}{max}\PY{l+s+s2}{\PYZdq{}}\PY{p}{, }\PY{n}{solver}\PY{o}{=}\PY{l+s+s2}{\PYZdq{}}\PY{l+s+s2}{cbc}\PY{l+s+s2}{\PYZdq{}}\PY{p}{}\PY{p}{)}
    \PY{n}{result\PYZus{}labels} \PY{o}{=} \PY{n}{milp\PYZus{}exact\PYZus{}labels}\PY{o}{.}\PY{n}{solve}\PY{p}{(}
        \PY{n}{rf}\PY{p}{, }\PY{l+m+mf}{0.3}\PY{p}{,}\PY{n}{labels}\PY{o}{=}\PY{p}{[}\PY{l+s+s2}{\PYZdq{}}\PY{l+s+s2}{blue}\PY{l+s+s2}{\PYZdq{}}\PY{p}{,}\PY{l+s+s2}{\PYZdq{}}\PY{l+s+s2}{green}\PY{l+s+s2}{\PYZdq{}}\PY{p}{,}\PY{l+s+s2}{\PYZdq{}}\PY{l+s+s2}{brown}\PY{l+s+s2}{\PYZdq{}}\PY{p}{]}\PY{p}{)}
    \PY{n}{print\PYZus{}result}\PY{p}{(}\PY{n}{result\PYZus{}labels}\PY{p}{)}
\end{Verbatim}

    \begin{Verbatim}[commandchars=\\\{\},fontsize=\fontsize{8}{9}]
{\color{outcolor}Out:} subsys states: 5, value: 1.0

    \end{Verbatim}
In contrast to minimizing the number of states, now taking the entire lower branch is optimal for maximal reachability probabilities and threshold \(0.3\). The objective value of
this subsystem is $1$, as it only includes one of the labels.

    \begin{Verbatim}[commandchars=\\\{\},fontsize=\fontsize{8}{9}]
{\color{incolor}In:} \PY{n}{result\PYZus{}labels}\PY{o}{.}\PY{n}{subsystem}\PY{o}{.}\PY{n}{digraph}\PY{p}{(}\PY{p}{)}
\end{Verbatim}

\noindent{\fontsize{8}{9}{\texttt{\color{outcolor}Out:}}}
    
    \begin{center}
    \adjustimage{max size={0.9\linewidth}{0.3\paperheight}}{running_example/mdp_sub_123.pdf}
    \end{center}

\section{Experiments}
\label{sec:experiments}
\newcommand{\psec}[1]{\nprounddigits{1}\npfourdigitnosep\numprint[s]{#1}}
\newcommand{\pnodes}[1]{\nprounddigits{0}\numprint{#1}}

We have run experiments on a number of models available in the benchmark suite\footnote{\url{https://github.com/prismmodelchecker/prism-benchmarks/}} of \prism.
The results and all scripts used to produce them are included in the supplementary material~\cite{FMCAD20-supplementary}. 
We used a computer with two Intel Xeon L5630 CPUs at 2.13GHz with four cores each and 189GB of RAM.
Each computation was assigned four cores, a memory limit of 10GB and each call to an LP/MILP solver (we use Gurobi, version 9.0.1) was limited to 20 minutes.

\begin{figure*}[t]
  \begin{subfigure}[t]{1\columnwidth}
  \begin{center}
    \resizebox{1.1\linewidth}{5.5cm}{
    \includegraphics{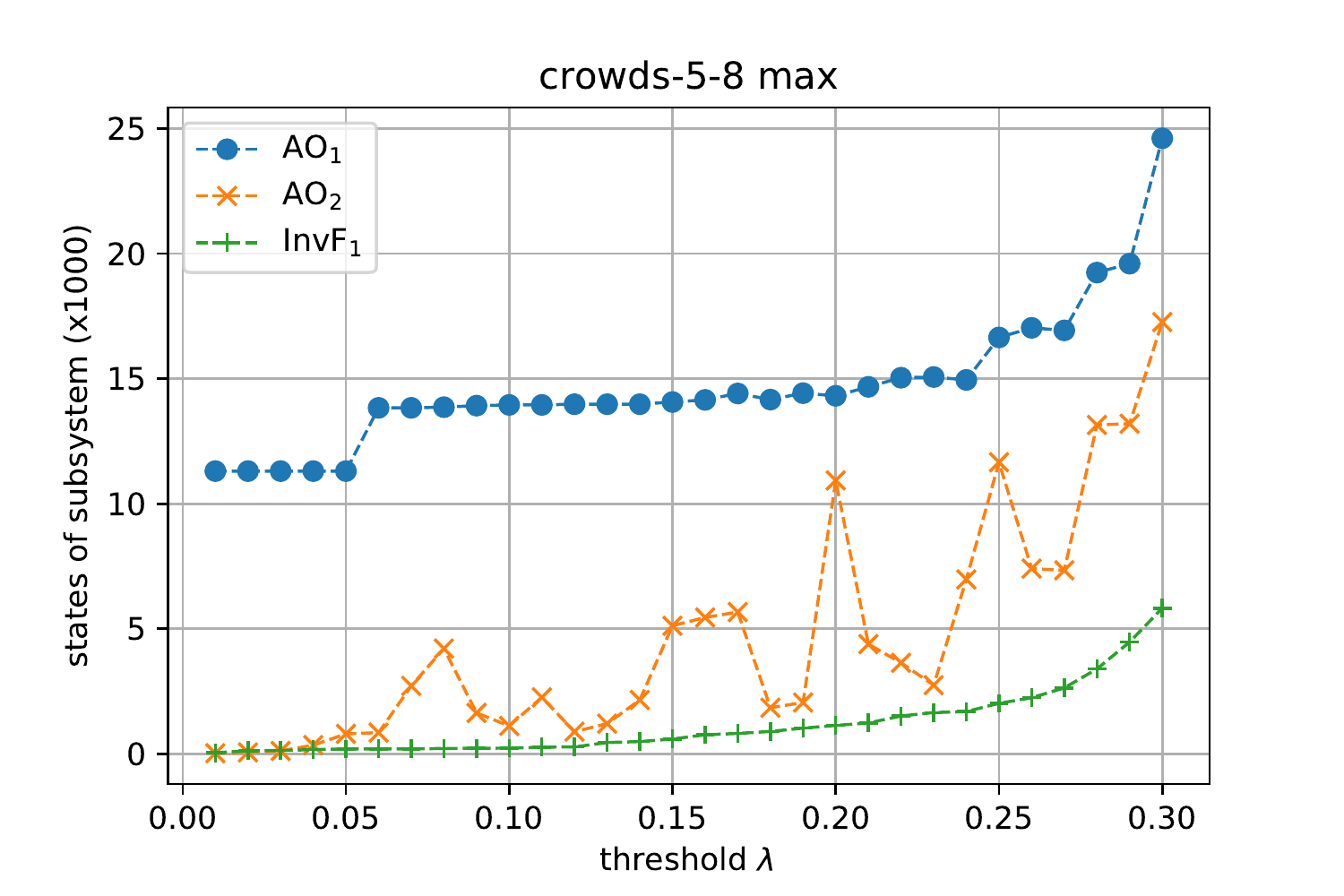}}
  \end{center}
  \label{subfig:crowds-5-8-max}
  \end{subfigure}
  \begin{subfigure}[t]{1\columnwidth}
  \begin{center}
    \resizebox{1.1\linewidth}{5.5cm}{
    \includegraphics{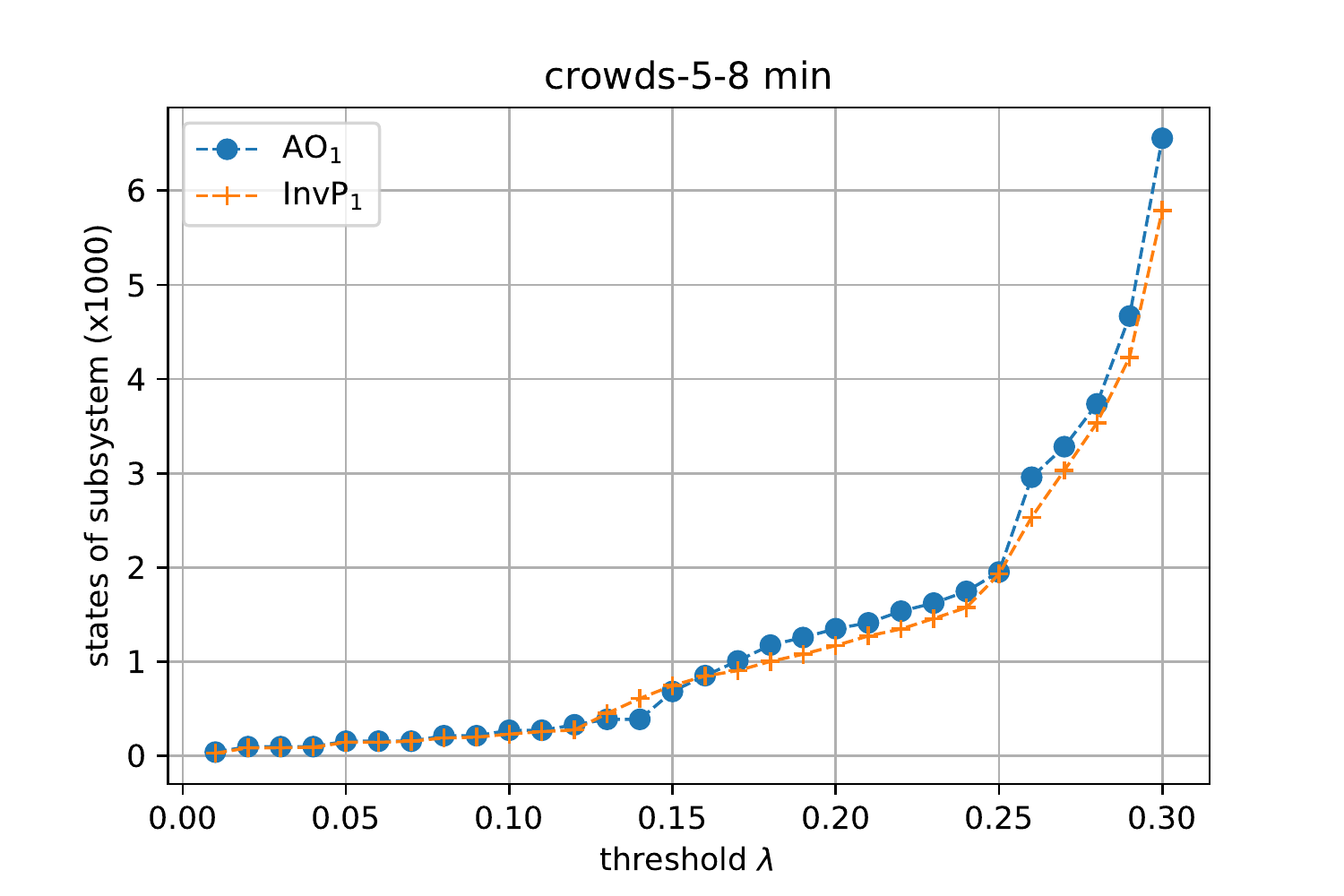}}
  \end{center}
  \label{subfig:crowds-5-8-min}
  \end{subfigure}

  \begin{subfigure}[t]{1\columnwidth}
  \begin{center}
    \resizebox{1.1\linewidth}{5.5cm}{
    \includegraphics{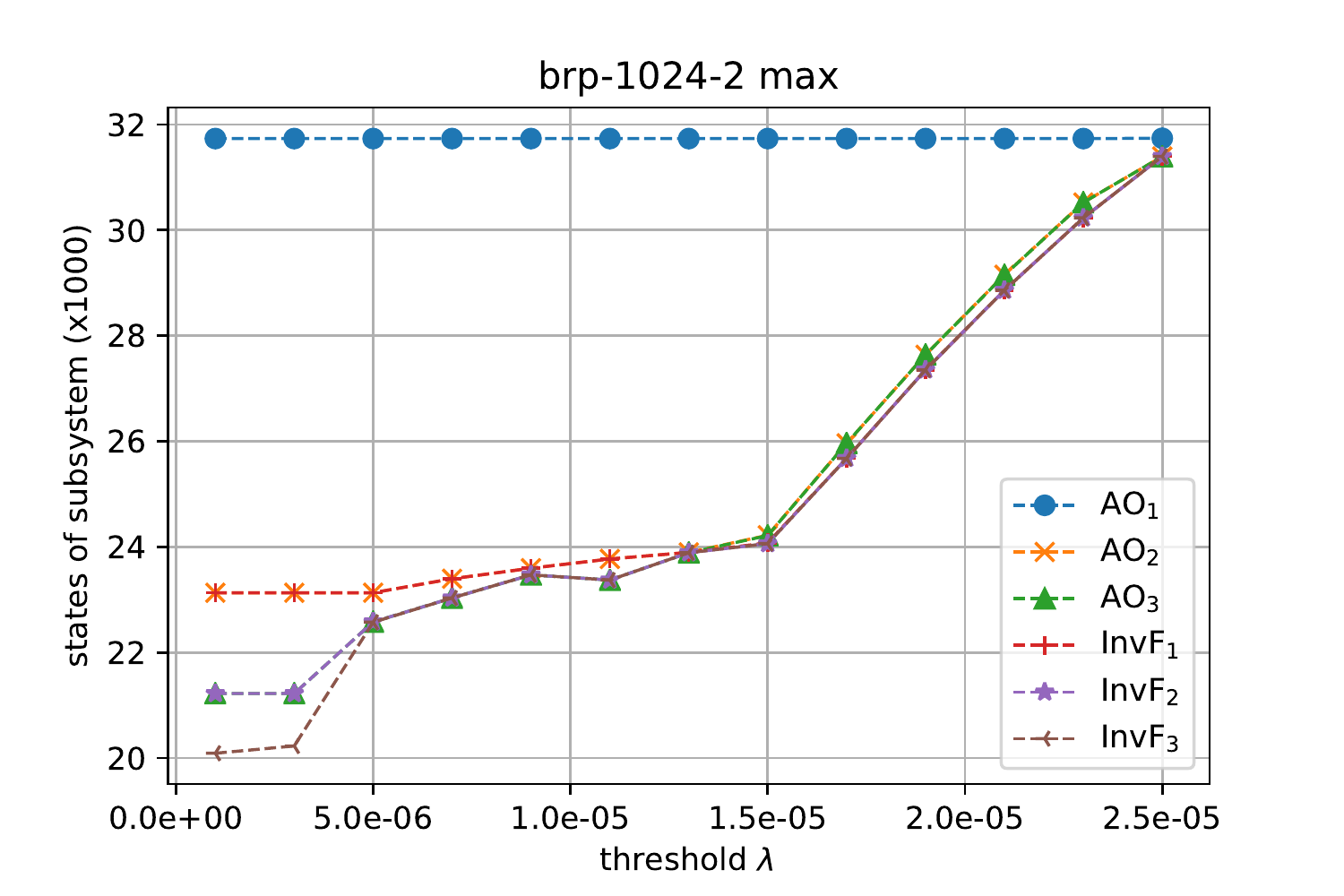}}
  \end{center}
  \label{subfig:brp-1024-2-max}
  \end{subfigure}
  \begin{subfigure}[t]{1\columnwidth}
  \begin{center}
    \resizebox{1.1\linewidth}{5.5cm}{
    \includegraphics{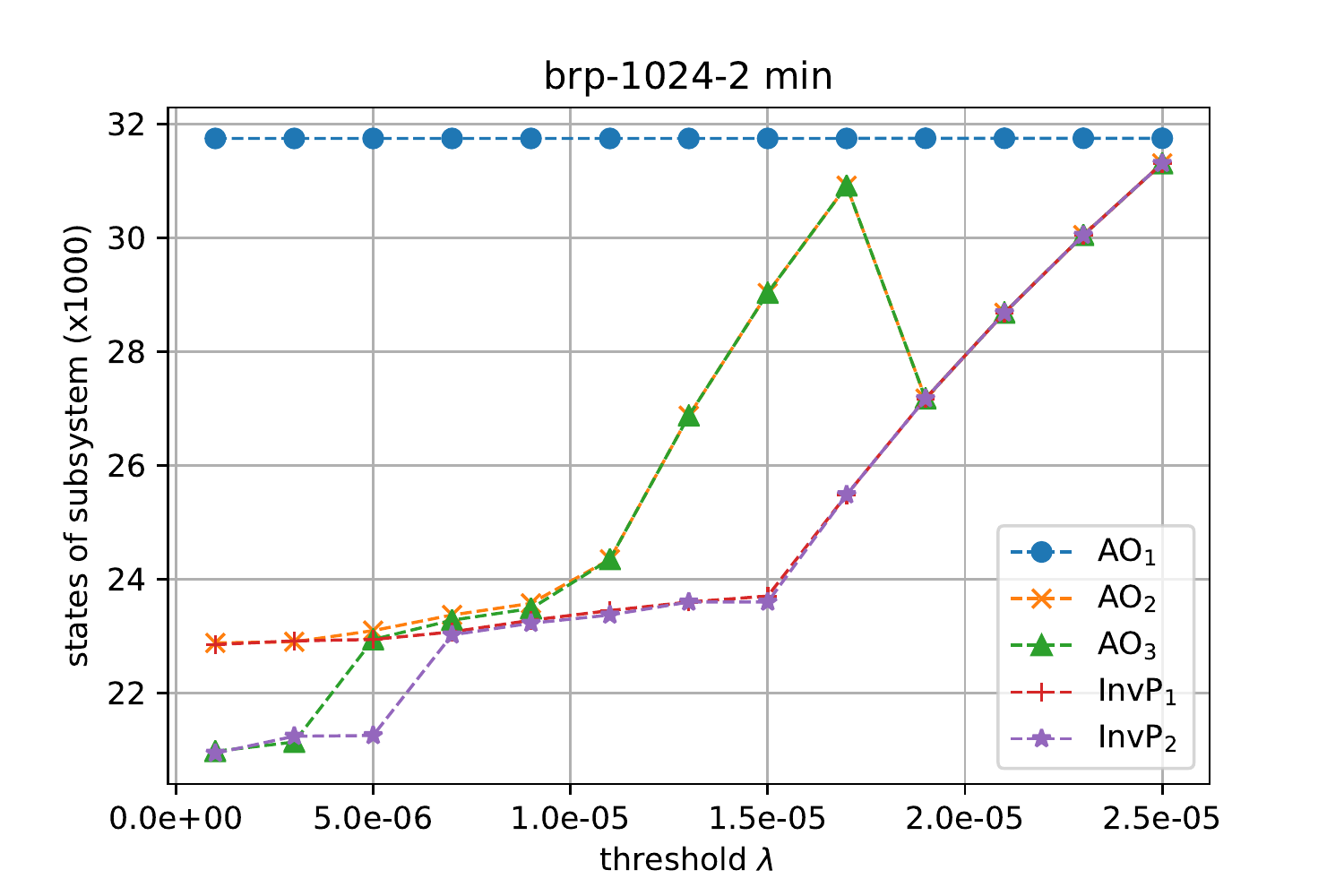}}
  \end{center}
  \label{subfig:brp-1024-2-min}
  \end{subfigure}
\caption{Comparison of subsystem sizes achieved by the heuristics of \toolname{} for benchmark DTMCs.}
 \label{fig:plots dtmc}
\end{figure*}

We consider the following models, where the first two are DTMCs and the last two are MDPs: the bounded retransmission protocol~\cite{DArgenioJJL01,HelminkSV93} (\texttt{brp-N-K}), the crowds protocol~\cite{Shmatikov04,ReiterR98} (\texttt{crowds-N-K}), the randomized consensus protocol~\cite{KwiatkowskaNS01,AspnesH90} (\texttt{consensus-N-K}) and the \texttt{csma-N-K} protocol for data channels~\cite{KwiatkowskaNSW07}.
In all cases increasing \texttt{N} and \texttt{K}, (which, for example, stand for the number of participating members, or a bound on possible random walks) leads to larger models.
For each model we fixed a reachability objective, inspired by properties considered in the benchmark suite.

\begin{figure*}[h]
	\begin{subfigure}[t]{1\columnwidth}
		\begin{center}
			\resizebox{1.1\linewidth}{5.5cm}{
				\includegraphics{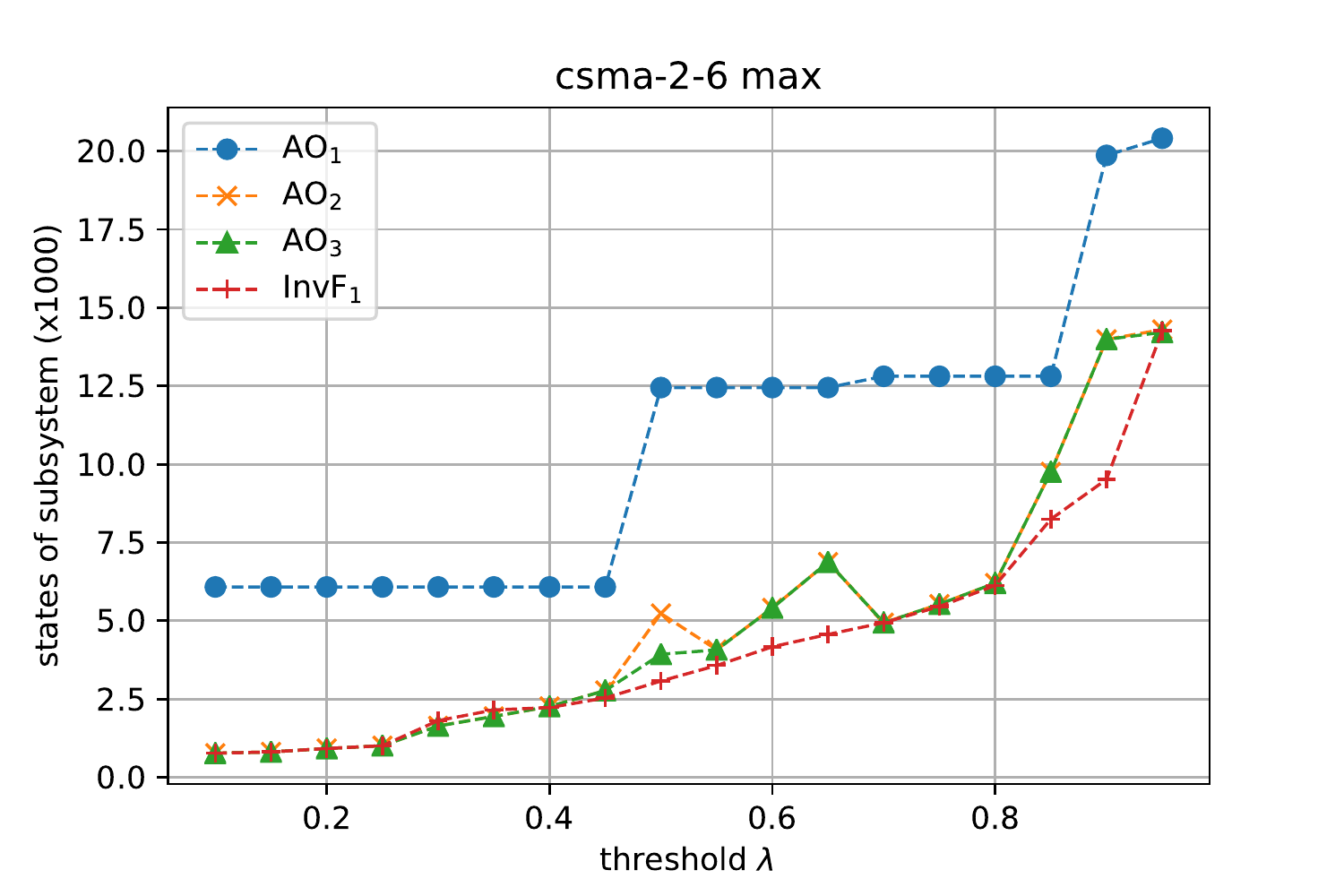}}
		\end{center}
		\label{subfig:csma-2-6-max}
	\end{subfigure}
	\begin{subfigure}[t]{1\columnwidth}
		\begin{center}
			\resizebox{1.1\linewidth}{5.5cm}{
				\includegraphics{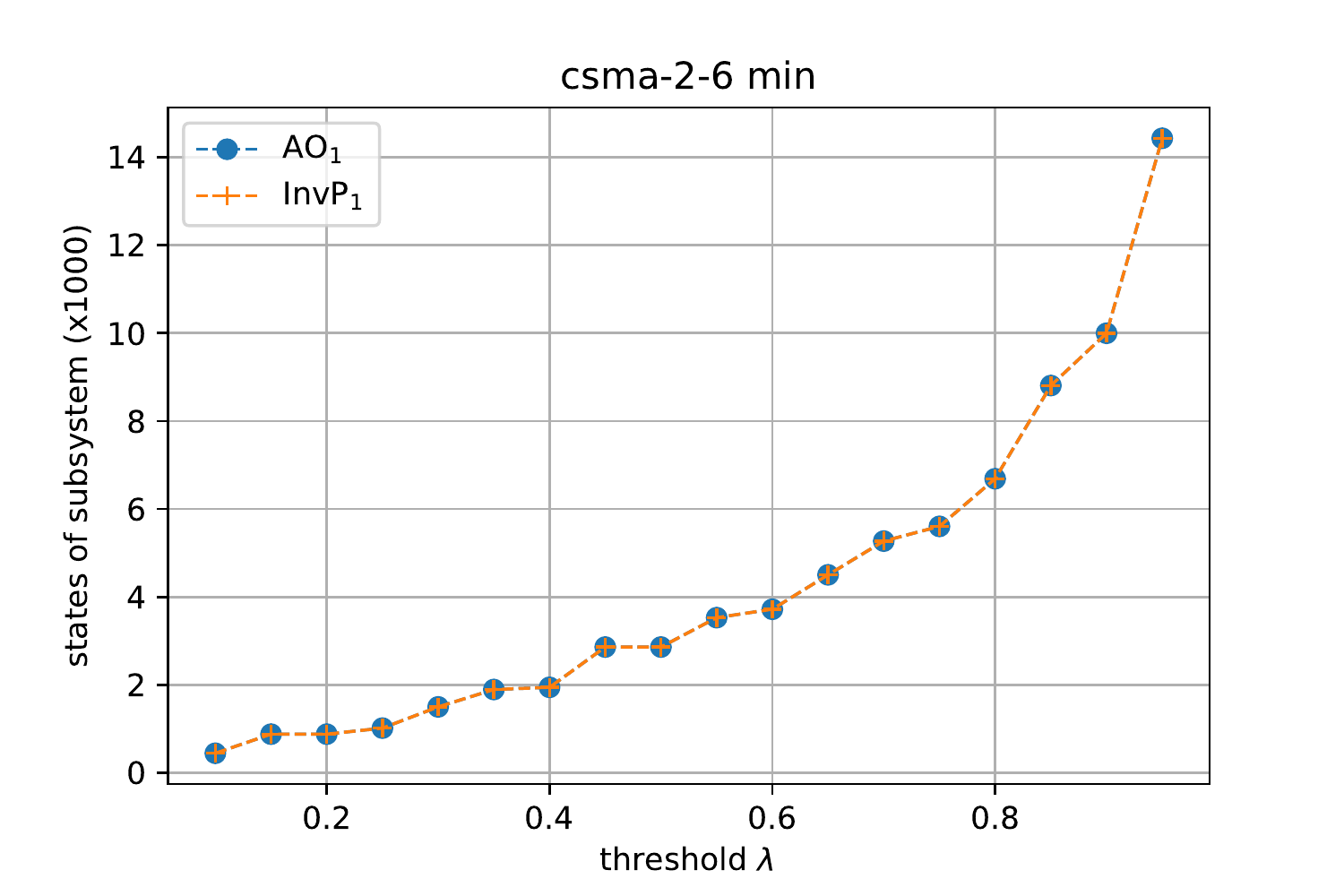}}
		\end{center}
		\label{subfig:csma-2-6-min}
	\end{subfigure}

	\begin{subfigure}[t]{1\columnwidth}
		\begin{center}
			\resizebox{1.1\linewidth}{5.5cm}{
				\includegraphics{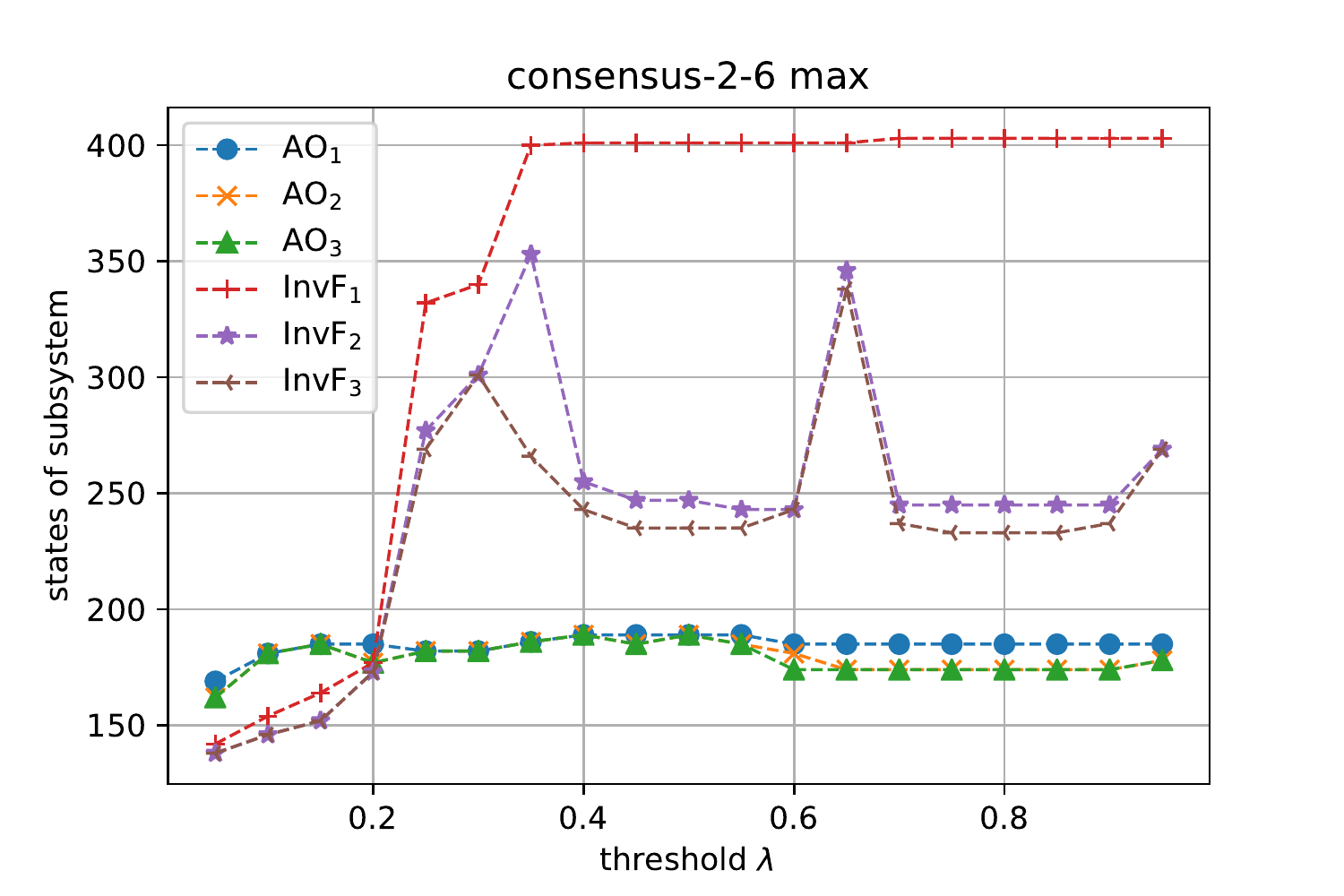}}
		\end{center}
		\label{subfig:consensus-2-6-max}
	\end{subfigure}
	\begin{subfigure}[t]{1\columnwidth}
		\begin{center}
			\resizebox{1.1\linewidth}{5.5cm}{
				\includegraphics{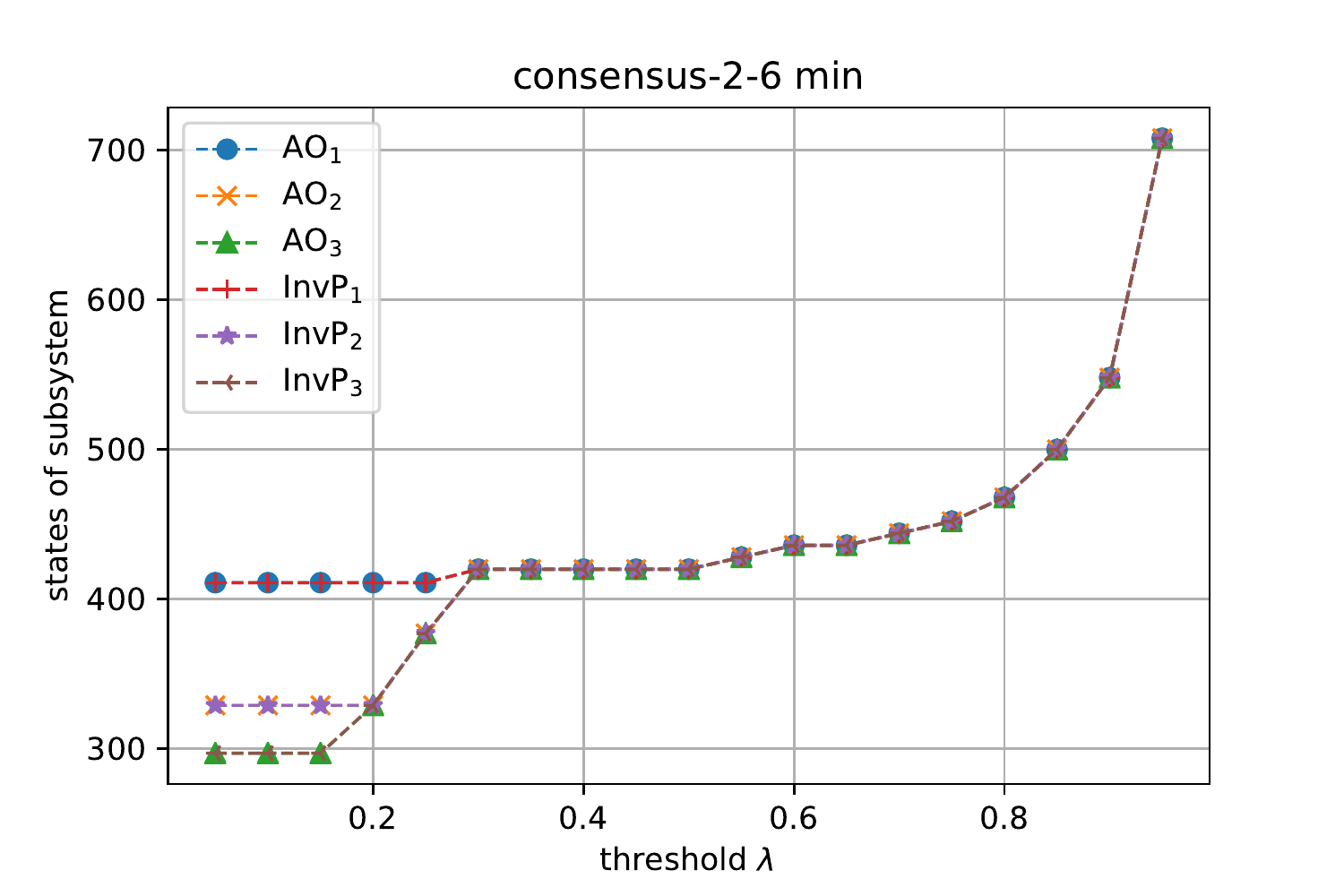}}
		\end{center}
		\label{subfig:consensus-2-6-min}
	\end{subfigure}
	\caption{Comparison of subsystem sizes achieved by the heuristics of \toolname{} for benchmark MDPs.}
	\label{fig:plots}
\end{figure*}

We contrast the results of the QS-heuristic with initial objective $(1,\ldots,1)$ (called AO for ``all-ones''), and the initial objectives $\invF{}$ and $\invP{}$, which are inverses of solutions of~\Cref{eqn:maxy,eqn:maxz}, respectively.
As $\invF{}$ is derived from the $\P^{\max}_{\M}(\lambda)$ polytope, we apply it to the max-queries, and conversely for $\invP{}$.
We let the QS-heuristic compute five iterations.
The subscripts $i$ in AO{$_i$}, $\invF{}_i$ and $\invP{}_i$ refer to the result at iteration $i$.
As the last iterations do not yield much improvement we only consider the first three iterations in~\Cref{fig:plots dtmc} and~\Cref{fig:plots}.
If no improvement was made after the $i$-th iteration, we do not show the following ones.

\newcommand{\mxtime}{\texttt{max-time}}
We examine for each model the time needed to compute the reachability form (from an explicit transition matrix) and the maximal time (over min/max-forms, all considered thresholds and initial values) needed to compute five iterations of the QS-heuristic, given the RF.
This latter value is called \mxtime{}.

We first consider the DTMCs: \texttt{crowds-5-8} (\pnodes{27849} states, \psec{11.345814047381282} to construct RF, \mxtime{}: \psec{191.62447188980877}) and \texttt{brp-1024-2} (\pnodes{31749} states, \psec{9.40612817183137} to construct RF, \mxtime{}: \psec{366.1451890747994}).
As $\prb^{\max}_{\M}(\lozenge \goal) = \prb^{\min}_{\M}(\lozenge \goal)$ if $\M$ is a DTMC, witnesses for max- and min-probabilities coincide.
Still, the QS-heuristic applied to the polytopes $\P^{\max}_{\M}(\lambda)$, and $\P^{\min}_{\M}(\lambda)$ yields different results.
This was already observed in~\cite{FunkeJB20}, where it was also noted that one of the two usually performs well with the initial objective AO (the only one considered in~\cite{FunkeJB20}).
The new experiments show that $\invF{}$ and $\invP{}$ are better initial vectors for the considered instances (see the difference between AO$_3$ and $\invF_1$ in \texttt{crowds-5-8 max} and, respectively, $\invP_2$ in \texttt{brp-1024-2 min} of \Cref{fig:plots dtmc}).
The new heuristics also tend to stabilise after fewer iterations.

The MDPs that we consider are: \texttt{consensus-2-6} (\pnodes{786} states, \pnodes{1170} state-action pairs, \psec{0.32598960027098656} to construct RF, \mxtime{}: \psec{3.5324833542108536}) and \texttt{CSMA-2-6} (\pnodes{66720} states, \pnodes{66790} state-action pairs, \psec{18.478221759200096} to construct RF, \mxtime{}: \psec{512.711826203391}).
The new heuristics have a mixed effect here: in the max-case, \texttt{CSMA-2-6} profits while for \texttt{consensus-2-6} the AO-initialization yields better results.
For min, AO and $\invP$ perform equally well.
It should be noted that in \texttt{CSMA-2-6} the number of actions per state is very close to one, and hence it is ``close'' to being a DTMC.
For \texttt{consensus-2-6} it is noteworthy that relatively small subsystems are possible for maximal reachability throughout all considered thresholds.

\begin{figure}[t]
	\begin{center}
		\resizebox{1.05\linewidth}{!}{
			\includegraphics{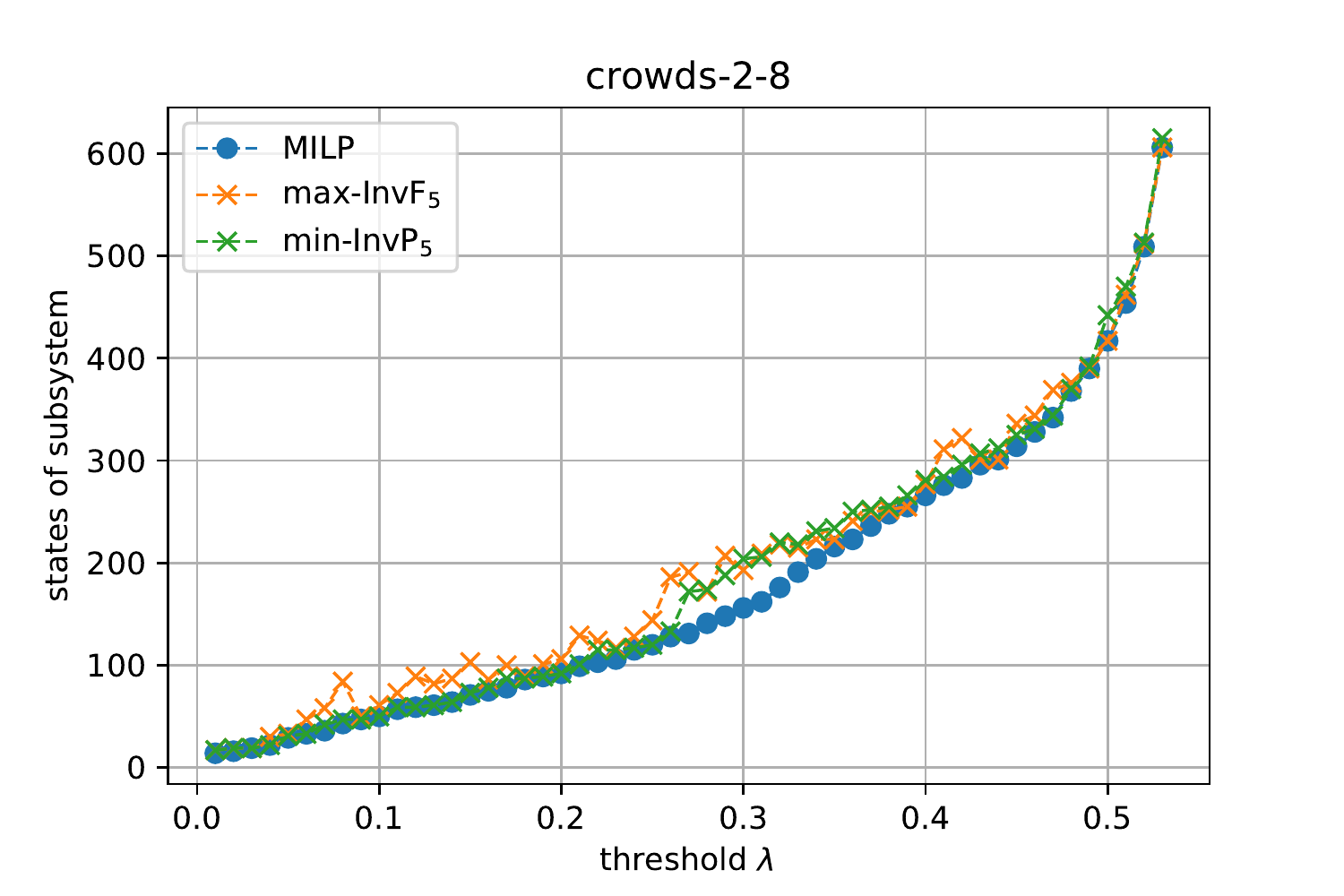}}
	\end{center}
	\caption{The QS-heuristic vs. exact minimization.}
	\label{fig:crowds-2-8-ex-vs-heur}
\end{figure}

The experiments show that the QS-heuristic is able to compute small witnessing subsystems in a reasonable time for models with over \pnodes{60000} states, and that the new heuristics perform well.
As the exact computations via MILP run into the timeout for all of the models in~\Cref{fig:plots dtmc} and~\Cref{fig:plots}, we cannot say how far the computed subsystems are from the optimal ones in terms of their size.
However, generalizing from smaller instances (see~\Cref{fig:crowds-2-8-ex-vs-heur}) indicates that the performance of the heuristic is good.

\section{Conclusion}

We have presented \toolname{}, a tool for computing small witnessing subsystems in discrete Markovian models. Contrary to other tools in the field, \toolname{} takes a unified approach for all scenarios that have been considered in the literature (minimal and maximal probabilities, exact and heuristic computation). New initial objective functions in the QS-heuristic have been shown to improve previous results for DTMCs. Our tool also comes with the complete functionality of a certificate generator and verifier for reachability problems in MDPs.

In future work we will investigate which properties of a DTMC benefit either the minimal or the maximal probability formulation, and add an automated detection scheme to \toolname{} in order to avoid redundant computations. We also intend to incorporate a new class of heuristics based on vertex enumeration algorithms.

\bibliographystyle{splncs04}
\bibliography{lit}

\end{document}